\documentclass[showpacs,preprintnumbers,amsmath,amssymb]{revtex4}
\usepackage{graphicx}
\usepackage{dcolumn}
\usepackage{bm}

\begin{document}
\title{Physical Consequences of a Theory\\ with Dynamical Volume Element}

\author
{E. I. Guendelman  and A.  B. Kaganovich}
\address{Physics Department, Ben Gurion University of the Negev, Beer
Sheva 84105, Israel}

\date{\today}

\begin{abstract}
We survey motivation, basic ideas and physical consequences of a
theory where the underlying action involves terms both with the
usual volume element $\sqrt{-g}d^{4}x$ and with the new one $\Phi
d^{4}x={4!}d\varphi_{1}\wedge d\varphi_{2}\wedge
d\varphi_{3}\wedge d\varphi_{4}$. The latter may be interpreted as
the 4-form determined on the 4-D space-time manifold (not
necessary Riemannian). Regarding the scalar fields $\varphi_{a}\,
(a=1,...4)$ as new dynamical variables and proceeding in the first
order formalism we realize the so-called Two Measures Theory which
possesses a number of attractive features and provides very
interesting predictions.
\end{abstract}

\maketitle

\begin{center}
\textbf{Content}
\end{center}
1. Motivation: The volume measure of the space-time manifold and
its dynamical degrees of freedom. \\
2. Classical equations of motion. \\
3. Scalar field model I. \\
4. Fine tuning free transition to $\Lambda =0$ state. \\
5. Scalar field model II: spontaneously broken global scale
invariance.\\
5.1  Formulation of the model.\\
5.2  Early power law inflation.\\
5.3 Quintessential inflation type of scenario\\
5.4 A tiny cosmological constant without fine
tuning of dimensionfull parameters.\\
6. Early TMT cosmology with no fine tuning: absence of the initial
singularity of the
curvature.\\
6.1 General analysis and numerical solutions.\\
6.2 Analysis of the initial singularity.\\
7. Sign indefiniteness of the manifold volume measure
 as the origin of a phantom dark energy.\\
8. Including fermions into the model II.\\
8.1. Description of the model.\\
8.2 Fermionic matter in normal conditions: reproducing GR and
fine tuning free decoupling from the quintessence field.\\
8.3  Nonrelativistic neutrinos and dark energy\\
8.4 Prediction of strong gravity effect in high energy
physics experiments.\\
9. Dust in normal conditions and its decoupling from the dark
energy. \\

\section{Motivation: The volume measure of the space-time manifold and
its dynamical degrees of freedom}

One of the motivations for using an additional measure of
integration in the action principle is closely related to a
possible degeneracy of the metric. Solutions with degenerate
metric were a subject of a long-standing discussions starting
probably with the paper by Einstein and Rosen\cite{Einstein}. In
spite of some difficulty interpreting solutions with degenerate
metric in classical theory of gravitation, the prevailing view was
that they have physical meaning and must be included in the path
integral\cite{Hawking1979},\cite{D'Auria-Regge},\cite{Tseytlin1982}.
In the first order formulation of an appropriately extended
general relativity , solutions with $g(x)\equiv
\det(g_{\mu\nu})=0$ allow to describe changes of the space-time
topology\cite{Hawking1979},\cite{Horowitz}. Similar idea is
realized also in the Ashtekar's
variables\cite{Ashtekar-3-in-Jacobson},\cite{Jacobson-2-in-Jacobson}.
There are known also classical
solutions\cite{Dray-0}-\cite{Senovilla} with change of the
signature of the metric tensor. The space-time regions with
$g(x)=0$ can be treated  as having {\em 'metrical dimension'}
$D<4$ (using terminology by Tseytlin\cite{Tseytlin1982}).

The simplest solution with $g(x)=0$ is $g_{\mu\nu}=0$ while the
affine connection is arbitrary (or, in the Einstein-Cartan
formulation, the vierbein $e_{\mu}^a =0$ and $\omega_{\mu}^{ab}$
is arbitrary). Such  solutions have been studied by D'Auria and
Regge\cite{D'Auria-Regge}, Tseytlin\cite{Tseytlin1982}),
Witten\cite{Witten-15-and-16-in-Horowitz},
Horowitz\cite{Horowitz}, Giddings\cite{Giddings},
Ba\~{n}ados\cite{Banados}; it has been suggested that
$g_{\mu\nu}=0$ should be interpreted as essentially non-classical
phase in which diffeomorphism invariance is unbroken and it is
realized at high temperature and curvature.

Now we would like to bring up a question: whether the equality
$g(x)=0$ really with a necessity means that the dimension of the
space-time manifold in a small neighborhood of the point $x$ may
become $D<4$? At first sight it should be so because the volume
element is
\begin{equation}
 dV_{(metrical)}=\sqrt{-g}d^{4}x.
 \label{dVg}
\end{equation}
 Note that the latter is the "metrical" volume
 element, and the possibility to describe the volume of the space-time manifold
 in this way appears after the 4-dimensional differentiable manifold $M_4$ is equipped with the
  metric structure. For a solution with $g_{\mu\nu} =0$, the situation with description of the space-time
   becomes even worse .
    However, in spite of lack of the metric, the manifold
 $M_4$ may still  possess a nonzero volume element and have the dimension $D=4$. The well
  known way to realize it consists
 in the construction of a differential 4-form build for example
 by means of four differential 1-forms $d\varphi_a$, ($a=1,2,3,4$): \,
 $d\varphi_1\wedge d\varphi_2\wedge d\varphi_3\wedge d\varphi_4$.
 Each of the 1-forms $d\varphi_a$ may be defined by a scalar field
  $\varphi_a(x)$. The appropriate volume element of the 4-dimensional
  differentiable manifold $M_4$
 can be represented in the following way
\begin{equation}
dV_{(manifold)} = {4!}d\varphi_{1}\wedge d\varphi_{2}\wedge
d\varphi_{3}\wedge d\varphi_{4}\equiv \Phi d^{4}x \label{dV}
\end{equation}
where
\begin{equation}
\Phi \equiv \varepsilon_{abcd} \varepsilon^{\mu\nu\lambda\sigma}
(\partial_{\mu}\varphi_{a}) (\partial_{\nu}\varphi_{b})
(\partial_{\lambda}\varphi_{c}) (\partial_{\sigma}\varphi_{d}).
\label{Phi}
\end{equation}
is the volume measure independent of $g_{\mu\nu}$ as opposed to
the case of the metrical volume measure $\sqrt{-g}$. In order to
emphasize the fact that the volume element (\ref{dV}) is metric
independent we will call it a {\em manifold volume element} and
the measure $\Phi$ - a {\em manifold volume measure}.

If $\Phi(x)\neq 0$ one can think of four scalar fields
$\varphi_a(x)$ as describing a  homeomorphism of an open
neighborhood of the point $x$ on the 4-dimensional Euclidean space
$R^4$. However if one allows a dynamical mechanism of metrical
dimensional reduction of the space-time by means of degeneracy of
the metrical volume measure $\sqrt{-g}$, there is no reason to
ignore a possibility of a similar effect permitting degenerate
manifold volume measure $\Phi$. The possibility of such (or even
stronger, with a sign change of $\Phi$) dynamical effect seems to
be here more natural since {\em the manifold volume measure $\Phi$
is sign indefinite} (in  Measure Theory, sign indefinite measures
are known as signed measures\cite{signed}) . Note that the
metrical and manifold volume measures are not obliged generically
to be simultaneously nonzero.

The original idea to use differential forms as describing
dynamical degrees of freedom of the space-time differentiable
manifold has been developed by Taylor in his attempt\cite{Taylor}
to quantize the gravity. Taylor argued that quantum mechanics is
not compatible with a Riemannian metric space-time; moreover, in
the quantum regime space-time is not even an affine manifold. Only
in the classical limit the metric and connection emerge, that one
allows then to construct a traditional space-time description. Of
course, the transition to the classical limit is described  in
Ref.\cite{Taylor} rather in the form of a general prescription.
Thereupon we would like to pay attention to the additional
possibility which was ignored in Ref.\cite{Taylor}. Namely, in the
classical limit not only the metric and connection emerge but also
some of the differential forms  could keep (or restore) certain
dynamical effect in the classical limit. In such a case, the
traditional space-time description may occur to be incomplete.
{\em Our key idea} is that one of such lost differential forms,
the 4-form (\ref{dV}), survives in the classical limit as
describing dynamical degrees of freedom of the volume measure of
the space-time manifold, and hence can affect the gravity theory
on the classical level too.

If we add four scalar fields $\varphi_a(x)$ as new  variables to a
set of usual variables  (like metric, connection and matter
degrees of freedom) which undergo variations in the action
principle then one can expect an effect of gravity and matter on
the manifold volume measure $\Phi$ and vice versa.

As is well known, the 4-dimensional differentiable manifold is
 orientable if it possesses a differential form  of degree 4
which is nonzero at every point on the manifold. Therefore two
possible signs of the manifold volume measure (\ref{Phi}) are
associated with two possible orientations of the space-time
manifold.
 The latter means that besides a
dimensional reduction and topology changes on the level of the
differentiable manifold, the incorporation of the manifold volume
measure $\Phi$ allows to realize solutions describing dynamical
change of the orientation of the space-time manifold.

 The simplest way
 to take into account the existence of two volume measures
consists in  the modification of the action which should now
consist of two terms, one with the usual measure $\sqrt{-g}$ and
another - with the measure $\Phi$,
\begin{equation}
    S_{mod} = \int \left(\Phi L_1 +\sqrt{-g}L_2\right) d^{4}x,
\label{sec-2-S-modif}
\end{equation}
where two Lagrangians $L_1$ and $L_2$ coupled with manifold and
metrical volume measures appear respectively.  According to our
previous experience\cite{GK1}-\cite{GK11} in Two Measures Field
Theory (TMT) we  proceed with an additional basic assumption that,
at least on the classical level, the Lagrangians $L_1$ and $L_2$
are independent of the scalar fields $\varphi_a(x)$, i.e. the
manifold volume measure degrees of freedom enter into TMT only
through the manifold volume measure $\Phi$.  In such a case,
  the action (\ref{sec-2-S-modif}) possesses
an infinite dimensional symmetry
\begin{equation}
\varphi_{a}\rightarrow\varphi_{a}+f_{a}(L_{1}), \label{IDS}
\end{equation}
 where
$f_{a}(L_{1})$ are arbitrary functions of  $L_{1}$ (see details in
Ref.\cite{GK3}). One can hope that this symmetry should prevent
emergence of the scalar fields $\varphi_a(x)$ dependence in
$L_{1}$ and $L_{2}$ after quantum effects are taken into account.

Note that Eq.(\ref{sec-2-S-modif}) is just a convenient way for
presentation of the theory in a general form. In concrete models,
 the action
(\ref{sec-2-S-modif}) can be always rewritten in an equivalent
form where {\em each term in the action has its own  total volume
measure and the latter is a linear combination of $\Phi$ and
$\sqrt{-g}$}.

\section{Classical equations of motion}
 Varying  $\varphi_{a}$'s, we get
$B^{\mu}_{a}\partial_{\mu}L_{1}=0$ where
$B^{\mu}_{a}=\varepsilon^{\mu\nu\alpha\beta}\varepsilon_{abcd}
\partial_{\nu}\varphi_{b}\partial_{\alpha}\varphi_{c}
\partial_{\beta}\varphi_{d}$.
Since $Det (B^{\mu}_{a}) = \frac{4^{-4}}{4!}\Phi^{3}$ it follows
that if $\Phi\neq 0$ the constraint
\begin{equation}
 L_{1}=sM^{4} =const.
\label{varphi}
\end{equation}
must be satisfied, where $s=\pm 1$ and $M$ is a constant of
integration with the dimension of mass. Variation of the metric
$g^{\mu\nu}$ gives
\begin{equation}
\zeta\frac{\partial L_1}{\partial g^{\mu\nu}}+\frac{\partial
L_2}{\partial g^{\mu\nu}}-\frac{1}{2}g_{\mu\nu}L_2 =0,
\label{g-mu-nu-varying}
\end{equation}
where
\begin{equation}
 \zeta\equiv \frac{\Phi}{\sqrt{-g}}
 \label{zeta}
\end{equation}
 is the scalar field
build of the scalar densities $\Phi$ and $\sqrt{-g}$.

We study  models with  the Lagrangians of the form
\begin{equation}
L_1=-\frac{1}{b_g\kappa}R(\Gamma, g)+L_1^m, \quad
L_2=-\frac{1}{\kappa}R(\Gamma, g)+L_2^m \label{L1L2}
\end{equation}
 where $\Gamma$ stands
for affine connection, $R(\Gamma,
g)=g^{\mu\nu}R_{\mu\nu}(\Gamma)$,
$R_{\mu\nu}(\Gamma)=R^{\lambda}_{\mu\nu\lambda}(\Gamma)$ and
$R^{\lambda}_{\mu\nu\sigma}(\Gamma)\equiv \Gamma^{\lambda}_{\mu\nu
,\sigma}+ \Gamma^{\lambda}_{\alpha\sigma}\Gamma^{\alpha}_{\mu\nu}-
(\nu\leftrightarrow\sigma)$. Dimensionless factor $b_g^{-1}$ in
front of $R(\Gamma, g)$ in $L_1$ appears because there is no
reason for couplings  of the scalar curvature to the measures
$\Phi$ and $\sqrt{-g}$ to be equal. We choose $b_g>0$ and $\kappa
=16\pi G$, $G$ is the Newton constant. $L_1^m$ and $L_2^m$ are the
matter Lagrangians which can include all possible terms used in
regular (with only volume measure $\sqrt{-g}$) field theory
models.

Since the measure $\Phi$ is sign indefinite, the total volume
measure $(\Phi/b_g +\sqrt{-g})$ in the gravitational term
$-\kappa^{-1}\int R(\Gamma, g)(\Phi/b_g +\sqrt{-g})d^4x$  is
generically also sign indefinite.

Variation of the connection yields the equations we have solved
earlier\cite{GK3}. The result is
\begin{equation}
\Gamma^{\lambda}_{\mu\nu}=\{
^{\lambda}_{\mu\nu}\}+\frac{1}{2}(\delta^{\alpha}_{\mu}\sigma,_{\nu}
+\delta^{\alpha}_{\nu}\sigma,_{\mu}-
\sigma,_{\beta}g_{\mu\nu}g^{\alpha\beta}) \label{GAM2}
\end{equation}
where $\{ ^{\lambda}_{\mu\nu}\}$  are the Christoffel's connection
coefficients of the metric $g_{\mu\nu}$ and $\sigma,_{\mu}\equiv
\zeta,_{\mu}/(\zeta +b_g)$.

If $\zeta\neq const.$ the covariant derivative of $g_{\mu\nu}$
with this connection is nonzero (nonmetricity) and consequently
geometry of the space-time with the metric $g_{\mu\nu}$ is
generically non-Riemannian.  The gravity and matter field
equations obtained by means of the first order formalism contain
both $\zeta$ and its gradient as well. It turns out that at least
at the classical level, the measure fields $\varphi_a$ affect the
theory only through the scalar field $\zeta$.

For the class of models (\ref{L1L2}), the consistency of the
constraint (\ref{varphi}) and the gravitational equations
(\ref{g-mu-nu-varying}) has the form of the following  constraint
\begin{equation}
 (\zeta
-b_g)(sM^4-L_1^m)+g^{\mu\nu}\left(\zeta \frac{\partial
L_{1m}}{\partial g^{\mu\nu}}+\frac{\partial L_2^m}{\partial
g^{\mu\nu}}\right)-2L_2^m=0, \label{Constr-original}
\end{equation}
which determines  $\zeta(x)$ (up to the chosen value of the
integration constant $sM^4$) as a local function of matter fields
and metric. Note that the geometrical object $\zeta(x)$ does not
have  its own dynamical equation of motion and its space-time
behavior is totally determined by  the metric and matter fields
dynamics via the constraint (\ref{Constr-original}). Together with
this, since $\zeta$ enters into all equations of motion, it
generically has straightforward effects on  dynamics of the matter
and gravity through the forms of potentials, variable fermion
masses and selfinteractions\cite{GK1}-\cite{GK11}.

For understanding the structure of TMT it is important to  note
that TMT (where, as we suppose, the scalar fields $\varphi_a$
enter only via the measure $\Phi$) is a constrained dynamical
system. In fact, the volume measure $\Phi$ depends only upon the
first derivatives of fields $\varphi_a$ and this dependence is
linear.  The  fields $\varphi_a$ do not have their own dynamical
equations: they are auxiliary fields.  All their dynamical effect
is displayed only in the following two ways:   a) in the
appearance of the scalar field $\zeta$ and its gradient in all
equations of motion; b) in generating the algebraic constraint
(\ref{varphi}) (or (\ref{Constr-original})) which determines
$\zeta$ as a function of matter fields and the metric.

\section{Scalar Field Model I}

Let us now study a model including gravity as in Eqs.(\ref{L1L2})
and a scalar field $\phi$. The action has the same structure as in
Eq.(\ref{sec-2-S-modif}) but  it is more convenient to write down
it in the form
\begin{eqnarray}
 S&=&S_g+S_{\phi}\quad \text{where}\label{S-model-scalar.f.}\\
S_g&=&-\frac{1}{b_g\kappa}\int d^4x (\Phi
+b_g\sqrt{-g})R(\Gamma,g)
\nonumber\\
 S_{\phi}&=&\frac{1}{b_g}\int d^4x
\left[(\Phi
+b_{\phi}\sqrt{-g})\frac{1}{2}g^{\mu\nu}\phi_{,\mu}\phi_{,\nu}
 -\Phi V_1(\phi)-\sqrt{-g}\,V_2(\phi)\right]
\nonumber
\end{eqnarray}
The appearance of the dimensionless factor $b_{\phi}$ is explained
by the fact that without fine tuning it is impossible in general
to provide the same coupling of the $\phi$ kinetic term to the
measures $\Phi$ and $\sqrt{-g}$. $V_1(\phi)$ and $V_2(\phi)$ are
potential-like functions; we will see below that the physical
potential of the scalar $\phi$ is a complicated function of
$V_1(\phi)$ and $V_2(\phi)$.

The constraint (\ref{Constr-original}) reads now
\begin{equation}
(\zeta
-b_g)[sM^4+V_1(\phi)]+2V_2(\phi)+b_g\frac{\delta}{2}g^{\alpha\beta}\phi_{,\alpha}\phi_{,\beta}
=0, \label{Scalar-f-Constr-original}
\end{equation}
where $\delta =(b_g-b_{\phi})/b_g$. Since $\zeta\neq const.$ the
connection (\ref{GAM2}) differs from the connection of the metric
$g_{\mu\nu}$. Therefore the space-time with the metric
$g_{\mu\nu}$ is non-Riemannian. To see the physical meaning of the
model we perform a transition to a new metric
\begin{equation}
\tilde{g}_{\mu\nu}=(\zeta +b_{g})g_{\mu\nu}, \label{gmunuEin}
\end{equation}
where the connection $\Gamma^{\lambda}_{\mu\nu}$ becomes equal to
the Christoffel connection coefficients of the metric
$\tilde{g}_{\mu\nu}$ and the space-time turns into (pseudo)
Riemannian. This is why the set of dynamical variables using the
metric $\tilde{g}_{\mu\nu}$ we call the Einstein frame. One should
point out that {\it the transformation} (\ref{gmunuEin}) {\em is
not a conformal} one since $(\zeta +b_{g})$ is sign indefinite.
But $\tilde{g}_{\mu\nu}$ is a regular pseudo-Riemannian metric.
For the action (\ref{S-model-scalar.f.}), gravitational equations
(\ref{g-mu-nu-varying}) in the Einstein frame take canonical GR
form with the same $\kappa =16\pi G$
\begin{equation}
G_{\mu\nu}(\tilde{g}_{\alpha\beta})=\frac{\kappa}{2}T_{\mu\nu}^{eff}
\label{gef}
\end{equation}
Here  $G_{\mu\nu}(\tilde{g}_{\alpha\beta})$ is the Einstein tensor
in the Riemannian space-time with the metric $\tilde{g}_{\mu\nu}$
and the energy-momentum tensor reads
\begin{eqnarray}
T_{\mu\nu}^{eff}&=&\frac{\zeta +b_{\phi}}{\zeta +b_{g}}
\left(\phi_{,\mu}\phi_{,\nu}- \tilde{g}_{\mu\nu}X\right)
-\tilde{g}_{\mu\nu}\frac{b_{g}-b_{\phi}}{(\zeta +b_{g})} X
+\tilde{g}_{\mu\nu}V_{eff}(\phi;\zeta,M)
 \label{Tmn-2}
\end{eqnarray}
where
\begin{equation}
X\equiv\frac{1}{2}\tilde{g}^{\alpha\beta}\phi_{,\alpha}\phi_{,\beta}
\label{X}
\end{equation}
and the function $V_{eff}(\phi;\zeta,M)$ is defined as following:
\begin{equation}
V_{eff}(\phi;\zeta ,M)= \frac{b_g\left[sM^{4}+V_{1}(\phi)\right]
-V_{2}(\phi)}{(\zeta +b_{g})^{2}}. \label{Veff1}
\end{equation}

The scalar $\phi$ field equation following from
Eq.(\ref{S-model-scalar.f.}) and rewritten in the Einstein frame
reads
\begin{equation}
\frac{1}{\sqrt{-\tilde{g}}}\partial_{\mu}\left[\frac{\zeta
+b_{\phi}}{\zeta
+b_{g}}\sqrt{-\tilde{g}}\tilde{g}^{\mu\nu}\partial_{\nu}\phi\right]
+\frac{\zeta V_1^{\prime}+ V_2^{\prime}}{(\zeta +b_{g})^2}=0
 \label{phief}
\end{equation}

The scalar field $\zeta$ in Eqs.(\ref{Tmn-2})-(\ref{phief}) is
determined by means of the constraint (\ref{Constr-original})
which in the Einstein frame (\ref{gmunuEin}) takes the form
\begin{equation}
(\zeta -b_{g})[sM^{4}+V_1(\phi)]+ 2V_2(\phi)+\delta\cdot
b_{g}(\zeta +b_{g})X =0.\label{constraint2-1}
\end{equation}

\section{Fine tuning free transition to $\Lambda =0$ state}

It is interesting to see the role of the manifold volume measure
in the resolution of the CC problem. We accomplish this now in the
framework of the scalar field model I of previous section. The
$\zeta$-dependence of $V_{eff}(\phi;\zeta ,M)$, Eq.(\ref{Veff1}),
in the form of inverse square like $(\zeta +b_{g})^{-2}$ has a key
role in the resolution of the old CC problem in TMT. One can show
that if quantum corrections to the underlying action generate
nonminimal coupling like  $\propto R(\Gamma,g)\phi^2$ in both
$L_1$ and $L_2$, the general form of the $\zeta$-dependence of
$V_{eff}$ remains similar: $V_{eff}\propto (\zeta +f(\phi))^{-2}$,
where $f(\phi)$ is a function. The fact that only such type of
$\zeta$-dependence emerges in $V_{eff}(\phi;\zeta ,M)$, and a
$\zeta$-dependence is absent for example in the numerator of
$V_{eff}(\phi;\zeta ,M)$, is a direct result of our basic
assumption that $L_1$ and $L_2$ in the action
(\ref{sec-2-S-modif}) are independent of the manifold measure
fields $\varphi_a$.

Generically, in the action (\ref{S-model-scalar.f.}),
$b_{\phi}\neq b_g$ that yields a nonlinear kinetic term (i.e. the
$k$-essence type dynamics) in the Einstein frame. But for purposes
of this section it is enough to take a simplified model with
$b_{\phi}= b_g$ (which is in fact a fine tuning) since the
nonlinear kinetic term has no qualitative effect on the zero CC
problem. In such a case $\delta =0$. Solving the constraint
(\ref{constraint2-1}) for $\zeta$ and substituting into
Eqs.(\ref{Tmn-2})-(\ref{phief}) we obtain equations for
scalar-gravity system which can be described by the regular GR
effective action with the scalar field potential
\begin{equation}
 V_{eff}(\phi)=
\frac{(sM^{4}+V_{1}(\phi))^2}{4[b_g(sM^4+V_1(\phi))-V_{2}(\phi)]}.
\label{Veff3}
\end{equation}
For an arbitrary nonconstant function $V_1(\phi)$ there exist
infinitely many values of the integration constant $sM^4$ such
that  $V_{eff}(\phi)$ has the {\em absolute minimum} at some
$\phi=\phi_0$ with $V_{eff}(\phi_0)=0$ (provided
$b_g[sM^4+V_1(\phi)]-V_2(\phi)>0$). This effect takes place as
$sM^4+V_1(\phi_0)=0$ {\em without fine tuning of the parameters
and initial conditions}. Note that the choice of the scalar field
potential in the GR effective action in a form proportional to a
perfect square like emerging in Eq.(\ref{Veff3}) would mean a fine
tuning.

 For illustrative purpose  let us consider the model\cite{GK11}
with
\begin{equation}
V_1(\phi)=\frac{1}{2}\mu_1^2\phi^2, \qquad
V_2(\phi)=V_2^{(0)}+\frac{1}{2}\mu_2^2\phi^2. \label{V12model}
\end{equation}
 Recall that adding a
constant to $V_1$ does not effect equations of motion, while
$V_2^{(0)}$ absorbs the bare CC and all possible vacuum
contributions. We take negative integration constant, i.e. $s=-1$,
and the only restriction on the values of the integration constant
$M$ and the parameters is that denominator in (\ref{Veff3}) is
positive.

Consider spatially flat FRW universe with the metric in the
Einstein frame
\begin{equation}
\tilde{g}_{\mu\nu}=diag(1,-a^2,-a^2,-a^2), \label{FRW}
\end{equation}
 where $a=a(t)$ is the
scale factor. Each cosmological solution ends with the transition
to a $\Lambda =0$ state via damping oscillations of the scalar
field $\phi$ towards its  absolute minimum $\phi_0$. The
appropriate oscillatory regime in the  phase plane is presented in
Fig. 1.
\begin{figure}[htb]
\begin{center}
\includegraphics[width=10.0cm,height=6.0cm]{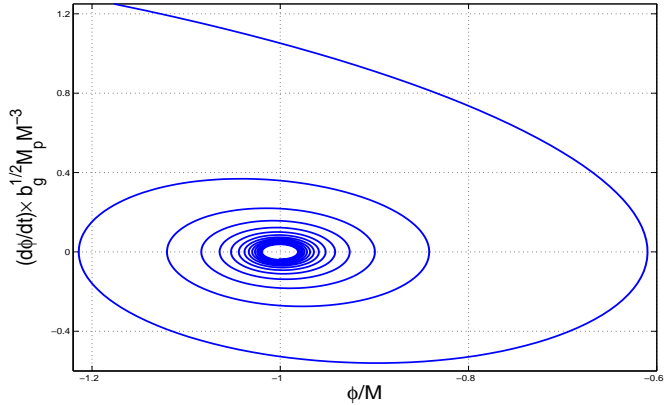}
\end{center}
 \caption{Typical phase curve (in the phase plane ($\phi$,$\dot{\phi})$)
  of the scalar field $\phi$ during
the transition to $\Lambda =0$ state. For illustrative purposes
the parameters are chosen such that
$V_{eff}=(M^2/2b_g)(\phi^2-M^2)^2/(\phi^2+4M^2)$ and $\phi_0=\pm
M$ and $\delta =0$. In the case without fine tuning of the
parameters $b_g\neq b_{\phi}$, i.e. $\delta\neq 0$, the phase
portrait is qualitatively the same.}\label{fig1}
\end{figure}

It follows from the constraint (\ref{constraint2-1}) (where  we
took $\delta =0$) that $|\zeta|\to\infty$ as $\phi\to\phi_0$. More
exactly, oscillations of $sM^{4}+V_1$ around zero are accompanied
with a singular behavior of $\zeta$ each time when $\phi$ crosses
$\phi_0$
\begin{equation}
\frac{1}{\zeta}\sim sM^{4}+V_1(\phi)\to 0 \qquad as \qquad
\phi\to\phi_0 \label{modul-zeta-infty}
\end{equation}
and $\zeta^{-1}$ oscillates around zero together with
$sM^{4}+V_1(\phi)$. Taking into account that the metric in the
Einstein frame $\tilde{g}_{\mu\nu}$, Eq.(\ref{FRW}), is regular we
deduce from Eq.(\ref{gmunuEin}) that the metric $g_{\mu\nu}$ used
in the underlying action (\ref{S-model-scalar.f.}) becomes
degenerate each time when $\phi$ crosses $\phi_0$
\begin{equation}
g_{00}=\frac{\tilde{g}_{00}}{\zeta +b_g}\sim \frac{1}{\zeta}\to 0;
\qquad g_{ii}=\frac{\tilde{g}_{ii}}{\zeta +b_g}\sim
-\frac{1}{\zeta}\to 0 \qquad as \qquad \phi\to\phi_0,
 \label{g00-degen}
\end{equation}
where we have taken into account that the energy density
approaches zero and therefore for this cosmological solution the
scale factor $a(t)$ remains finite in all times $t$. Therefore
\begin{equation}
\sqrt{-g}\sim \frac{1}{\zeta^2}\to 0 \qquad and \qquad \Phi
=\zeta\sqrt{-g}\sim \frac{1}{\zeta}\to 0\qquad as \qquad
\phi\to\phi_0\label{sqrtg-degen}
\end{equation}

 The detailed behavior of $\zeta$, the manifold
measure $\Phi$ and $g_{\mu\nu}$ - components\footnote{Since the
metric in the Einstein frame $\tilde{g}_{\mu\nu}$ is diagonal,
Eq.(\ref{FRW}), it is clear from the transformation
(\ref{gmunuEin}) that $g_{\mu\nu}$ is also diagonal.}  are shown
in Fig. 2.
\begin{figure}[htb]
\includegraphics[width=10.0cm,height=7.0cm]{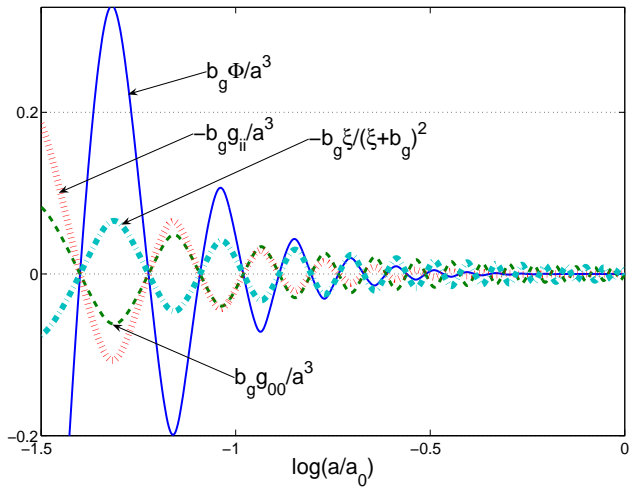}
\caption{Oscillations of the measure $\Phi$, the original metric
$g_{\mu\nu}$ and $\zeta/(\zeta +b_g)^2$ during the transition to
$\Lambda =0$ state. }\label{fig2}\end{figure}

Recall that the manifold volume measure $\Phi$ is a signed
measure\cite{signed} and therefore it  is not a surprise that it
can change sign. But TMT shows that including the manifold degrees
of freedom into the dynamics of the scalar-gravity system we
discover an interesting dynamical effect: a transition to zero
vacuum energy is accompanied by oscillations of $\Phi$ around
zero. Similar oscillations simultaneously occur with all
components of the metric $g_{\mu\nu}$ used in the underlying
action (\ref{S-model-scalar.f.}).

 The measure $\Phi$ and
the metric $g_{\mu\nu}$ pass zero  only in a discrete set of
moments in the course of  transition to the $\Lambda =0$ state.
Therefore there is no problem with the condition $\Phi\neq 0$ used
for the solution (\ref{varphi}). Also there is no problem with
singularity of $g^{\mu\nu}$ in the underlying action  since
\begin{equation}
\lim_{\phi\to\phi_0}\Phi g^{\mu\nu}=finite \qquad and \qquad
\sqrt{-g} g^{\mu\nu}\sim\frac{1}{\zeta}\to 0 \qquad as \qquad
\phi\to\phi_0 \label{lim}
\end{equation}
The metric in the Einstein frame $\tilde{g}_{\mu\nu}$ is always
regular because degeneracy of $g_{\mu\nu}$ is compensated in
Eq.(\ref{gmunuEin}) by singularity of the ratio of two measures
$\zeta\equiv\Phi/\sqrt{-g}$.

In Ref.(\cite{GK9}) we have explained in details how the presented
resolution of the old CC problem avoids the well known no-go
theorem by Weinberg\cite{Weinberg1} stating that generically in
field theory one cannot achieve zero value of the potential in the
minimum without fine tuning. It is interesting that the resolution
of the old CC problem in the context of TMT happens in the regime
where $\zeta\rightarrow\infty$. From the point of view of TMT, the
latter is the answer to the question why the old cosmological
constant problem cannot be solved (without fine tuning) in
theories with only the measure of integration $\sqrt{-g}$  in the
action.

\section{Scalar field model II. Spontaneously broken global scale invariance}
\subsection{Formulation of the model}

Let us now turn to the analyze of the results of the TMT model
possessing a global scale invariance studied early in
detail\cite{G1}-\cite{GKatz},\cite{GK5}-\cite{GK9}. The scalar
field $\phi$ playing the role of a model of dark energy appears
here as a dilaton, and a spontaneous breakdown of the scale
symmetry results directly from the presence of the manifold volume
measure $\Phi$. In other words, this SSB is an intrinsic feature
of TMT.

The action of the model reads
\begin{eqnarray}
 &&S=S_g+S_{\phi} \quad \text{where}\label{totaction}\\
S_g&=&-\frac{1}{b_g\kappa}\int d^{4}x e^{\alpha\phi /M_{p}}(\Phi
+b_{g}\sqrt{-g})
R(\Gamma ,g)\nonumber\\
S_{\phi}&=&\frac{1}{b_g}\int d^{4}x e^{\alpha\phi
/M_{p}}\left[(\Phi
+b_{\phi}\sqrt{-g})\frac{1}{2}g^{\mu\nu}\phi_{,\mu}\phi_{,\nu}-e^{\alpha\phi
/M_{p}}\left(\Phi V_{1} +\sqrt{-g}V_{2}\right)\right] \nonumber
\end{eqnarray}
 and it is invariant under the global scale transformations ($\theta =const.$):
\begin{equation}
    g_{\mu\nu}\rightarrow e^{\theta }g_{\mu\nu}, \,\,
\Gamma^{\mu}_{\alpha\beta}\rightarrow \Gamma^{\mu}_{\alpha\beta},
\,\, \varphi_{a}\rightarrow \lambda_{ab}\varphi_{b}\,\,
\text{where} \,\, \det(\lambda_{ab})=e^{2\theta}, \,\,
\phi\rightarrow \phi-\frac{M_{p}}{\alpha}\theta . \label{st}
\end{equation}
The appearance of the dimensionless parameters $b_g$ and
$b_{\phi}$ is explained by the same reasons we mentioned after
Eqs.(\ref{L1L2}) and (\ref{S-model-scalar.f.}). In contrast to the
model of Sec.3, now we deal with exponential (pre-) potentials
where $V_1$ and $V_2$ are constant dimensionfull parameters. The
remarkable feature of this TMT model is that Eq.(\ref{varphi}),
being the solution of the equation of motion resulting from
variation of the manifold volume measure degrees of freedom,
breaks spontaneously the scale symmetry (\ref{st}): this happens
due to the appearance of a dimensionfull integration constant
$sM^4$ in Eq.(\ref{varphi}).

One can show\cite{GK9} that if $b_{g}V_{1}> V_{2}$ then  the
ground state appears to be again (as it was in the scalar field
model I, see Secs.3 and 4) a zero CC state without fine tuning of
the parameters and initial conditions in the following two cases:
either $s=-1$ and $V_1>0$ or $s=+1$ and $V_1<0$. The behavior of
$\Phi$ and $g_{\mu\nu}$ in the course of transition to the
$\Lambda =0$ state is qualitatively the same as we observed in
Sec.4.

Here, in the presentation of the results of the model
(\ref{totaction}), we restrict ourself with the choice $s=+1$ and
$V_1<0$.

Similar to the model of Sec.3, equations of motion resulting from
the action (\ref{totaction}) are noncanonical and the space-time
is non Riemannian when using the original set of variables. This
is because  all the equations of motion and the solution for the
connection coefficients include terms proportional to
$\partial_{\mu}\zeta$. However, when working with the new metric
($\phi$
 remains the same)
\begin{equation}
\tilde{g}_{\mu\nu}=e^{\alpha\phi/M_{p}}(\zeta +b_{g})g_{\mu\nu},
\label{ct}
\end{equation}
which we call the Einstein frame,
 the connection  becomes Riemannian and  general form of all equations
 becomes canonical. Since
$\tilde{g}_{\mu\nu}$ is invariant under the scale transformations
(\ref{st}), spontaneous breaking of the scale symmetry  is reduced
in the Einstein frame to the {\it spontaneous breakdown of the
shift symmetry}
\begin{equation}
 \phi\rightarrow\phi +const.
 \label{phiconst}
\end{equation}
Notice that the Goldstone theorem generically is not applicable in
this theory\cite{G1}. The reason is the following. In fact, the
scale symmetry (\ref{phiconst}) leads to a conserved dilatation
current $j^{\mu}$. However, for example in the spatially flat FRW
universe the spatial components of the current $j^{i}$ behave as
$j^{i}\propto M^4x^i$ as $|x^i|\rightarrow\infty$. Due to this
anomalous behavior at infinity, there is a flux of the current
leaking to infinity, which causes the non conservation of the
dilatation charge. The absence of the latter implies that one of
the conditions necessary for the Goldstone theorem is missing.

 After the change of
variables  to the Einstein frame (\ref{ct}) the gravitational
equation takes the standard GR form with the same Newton constant
as in (\ref{totaction})
\begin{equation}
G_{\mu\nu}(\tilde{g}_{\alpha\beta})=\frac{\kappa}{2}T_{\mu\nu}^{eff}
 \label{gef}
\end{equation}
where  $G_{\mu\nu}(\tilde{g}_{\alpha\beta})$ is the Einstein
tensor in the Riemannian space-time with the metric
$\tilde{g}_{\mu\nu}$. The energy-momentum tensor
$T_{\mu\nu}^{eff}$ reads
\begin{eqnarray}
T_{\mu\nu}^{eff}&=&
 \frac{\zeta +b_{\phi}}{\zeta +b_{g}}
\left(\phi_{,\mu}\phi_{,\nu} -\frac{1}{2}
\tilde{g}_{\mu\nu}\tilde{g}^{\alpha\beta}\phi_{,\alpha}\phi_{,\beta}\right)
\label{Tmn}\\
&-&\tilde{g}_{\mu\nu}\frac{b_{g}-b_{\phi}}{2(\zeta +b_{g})}
\tilde{g}^{\alpha\beta}\phi_{,\alpha}\phi_{,\beta}
+\tilde{g}_{\mu\nu}V_{eff}(\phi,\zeta;M) \nonumber
\end{eqnarray}
where the function $V_{eff}(\phi,\zeta;M)$ is defined as
following:
\begin{equation}
V_{eff}(\phi,\zeta;M)=
\frac{b_{g}\left[M^{4}e^{-2\alpha\phi/M_{p}}+V_{1}\right]
-V_{2}}{(\zeta +b_{g})^{2}}. \label{Veff2}
\end{equation}

The scalar field $\zeta$  is determined by means of the constraint
similar to Eq.(\ref{constraint2-1}) of Sec.IV
\begin{eqnarray}
&&(b_{g}-\zeta)\left[M^{4}e^{-2\alpha\phi/M_{p}}+
V_{1}\right]-2V_{2}-\delta\cdot b_{g}(\zeta +b_{g})X
=0\label{constraint2}
\end{eqnarray}
where
\begin{equation}
X\equiv\frac{1}{2}\tilde{g}^{\alpha\beta}\phi_{,\alpha}\phi_{,\beta}
\qquad \text{and} \qquad \delta =\frac{b_{g}-b_{\phi}}{b_{g}}
\label{delta}
\end{equation}

The dilaton $\phi$ field equation in the Einstein frame is reduced
to the following
\begin{equation}
\frac{1}{\sqrt{-\tilde{g}}}\partial_{\mu}\left[\frac{\zeta
+b_{\phi}}{\zeta
+b_{g}}\sqrt{-\tilde{g}}\tilde{g}^{\mu\nu}\partial_{\nu}\phi\right]-\frac{2\alpha\zeta}{(\zeta
+b_{g})^{2}M_{p}}M^{4}e^{-2\alpha\phi/M_{p}} =0.
\label{phi-after-con}
\end{equation}
where again $\zeta$  is a solution of the constraint
(\ref{constraint2}). Note that the dilaton $\phi$ dependence in
all equations of motion in the Einstein frame appears {\em only}
in the form $M^{4}e^{-2\alpha\phi/M_{p}}$, i.e. it results only
from the spontaneous breakdown of the scale symmetry (\ref{st}).

The effective energy-momentum tensor (\ref{Tmn}) can be
represented in a form of that of  a perfect fluid
\begin{equation}
T_{\mu\nu}^{eff}=(\rho +p)u_{\mu}u_{\nu}-p\tilde{g}_{\mu\nu},
\qquad \text{where} \qquad
u_{\mu}=\frac{\phi_{,\mu}}{(2X)^{1/2}}\label{Tmnfluid}
\end{equation}
with the following energy and pressure densities resulting from
Eqs.(\ref{Tmn}) and (\ref{Veff2}) after inserting the solution
$\zeta =\zeta(\phi,X;M)$ of Eq.(\ref{constraint2}):
\begin{equation}
\rho(\phi,X;M) =X+ \frac{(M^{4}e^{-2\alpha\phi/M_{p}}+V_{1})^{2}-
2\delta b_{g}(M^{4}e^{-2\alpha\phi/M_{p}}+V_{1})X -3\delta^{2}
b_{g}^{2}X^2}{4[b_{g}(M^{4}e^{-2\alpha\phi/M_{p}}+V_{1})-V_{2}]},
\label{rho1}
\end{equation}
\begin{equation}
p(\phi,X;M) =X- \frac{\left(M^{4}e^{-2\alpha\phi/M_{p}}+V_{1}+
\delta b_{g}X\right)^2}
{4[b_{g}(M^{4}e^{-2\alpha\phi/M_{p}}+V_{1})-V_{2}]}. \label{p1}
\end{equation}

In a spatially flat FRW universe with the metric
$\tilde{g}_{\mu\nu}=diag(1,-a^2,-a^2,-a^2)$ filled with the
homogeneous scalar field $\phi(t)$, the $\phi$  field equation of
motion takes the form
\begin{equation}
Q_{1}\ddot{\phi}+ 3HQ_{2}\dot{\phi}- \frac{\alpha}{M_{p}}Q_{3}
M^{4}e^{-2\alpha\phi/M_{p}}=0 \label{phi1}
\end{equation}
 where $H$ is the Hubble parameter and we have used the notations
$\dot{\phi}\equiv d\phi/dt$,
\begin{equation}
Q_1=2[b_{g}(M^{4}e^{-2\alpha\phi/M_{p}}+V_{1})-V_{2}]\rho_{,X}
=(b_{g}+b_{\phi})(M^{4}e^{-2\alpha\phi/M_{p}}+V_{1})-
2V_{2}-3\delta^{2}b_{g}^{2}X \label{Q1}
\end{equation}
\begin{equation}
Q_2=2[b_{g}(M^{4}e^{-2\alpha\phi/M_{p}}+V_{1})-V_{2}]p_{,X}=
(b_{g}+b_{\phi})(M^{4}e^{-2\alpha\phi/M_{p}}+V_{1})-
2V_{2}-\delta^{2}b_{g}^{2}X\label{Q2}
\end{equation}
\begin{eqnarray}
\label{Q3} &&\quad
Q_{3}=\frac{1}{[b_{g}(M^{4}e^{-2\alpha\phi/M_{p}}+V_{1})-V_{2}]}
\\
&&\times \left[(M^{4}e^{-2\alpha\phi/M_{p}}+V_{1})
[b_{g}(M^{4}e^{-2\alpha\phi/M_{p}}+V_{1})-2V_{2}] +2\delta
b_{g}V_{2}X+3\delta^{2}b_{g}^{3}X^{2}\right] \nonumber
\end{eqnarray}
 The non-linear $X$-dependence appears here in the
framework of the fundamental theory without exotic terms in the
Lagrangians $L_1$ and $L_2$.  This effect follows just from the
fact that there are no reasons to choose the parameters $b_{g}$
and $b_{\phi}$ in the action (\ref{totaction}) to be equal in
general; on the contrary, the choice $b_{g}=b_{\phi}$ would be a
fine tuning.  Thus the above equations represent an  explicit
example of $k$-essence\cite{k-essence}  resulting from first
principles. The system of equations (\ref{gef}),
(\ref{rho1})-(\ref{phi1}) accompanied with the functions
(\ref{Q1})-(\ref{Q3}) and written in the metric
$\tilde{g}_{\mu\nu}=diag(1,-a^2,-a^2,-a^2)$ can be obtained from
the k-essence type effective action
\begin{equation}
S_{eff}=\int\sqrt{-\tilde{g}}d^{4}x\left[-\frac{1}{\kappa}R(\tilde{g})
+p\left(\phi,X;M\right)\right] \label{k-eff},
\end{equation}
where $p(\phi,X;M)$ is given by Eq.(\ref{p1}). In contrast to the
simplified models studied in literature\cite{k-essence}, it is
impossible here to represent $p\left(\phi,X;M\right)$ in a
factorizable form like $\tilde{K}(\phi)\tilde{p}(X)$. The scalar
field effective Lagrangian, Eq.(\ref{p1}), can be represented in
the form
\begin{equation}
p\left(\phi,X;M\right)=K(\phi)X+ L(\phi)X^2-U(\phi)
\label{eff-L-ala-Mukhanov}
\end{equation}
where the potential
\begin{equation}
U(\phi)=\frac{[V_{1}+M^{4}e^{-2\alpha\phi/M_{p}}]^{2}}
{4[b_{g}\left(V_{1}+M^{4}e^{-2\alpha\phi/M_{p}}\right)-V_{2}]}
 \label{eff-L-ala-Mukhanov-potential}
\end{equation}
and $K(\phi)$ and $L(\phi)$ depend on $\phi$ only via
$M^{4}e^{-2\alpha\phi/M_{p}}$. Notice that $U(\phi)>0$  for any
$\phi$ provided
\begin{equation}
b_g>0, \qquad  \text{and} \qquad b_{g}V_{1}\geq V_{2}
\label{bV1>V2},
\end{equation}
that we will assume in what follows.  Note  that besides the
presence of the effective potential $U(\phi)$, the Lagrangian
$p\left(\phi,X;M\right)$ differs from that of
Ref.\cite{k-inflation-Mukhanov} by the sign of $L(\phi)$: in our
case $L(\phi)<0$ provided the conditions (\ref{bV1>V2}). This
result cannot be removed by a choice of the parameters of the
underlying action (\ref{totaction}) while in
Ref.\cite{k-inflation-Mukhanov} the positivity of $L(\phi)$ was an
essential {\it assumption}. This difference plays a crucial role
for a possibility of a dynamical protection from the initial
singularity of the curvature studied in detail in Ref\cite{GK9}.

The model allows a power law inflation (where the dilaton $\phi$
plays the role of the inflaton) with a graceful exit to a zero or
tiny cosmological constant state.  In what it concerns to
primordial perturbations of $\phi$ and their evolution, there are
no difference with the usual (i.e. one-measure) model with the
action (\ref{k-eff})-(\ref{eff-L-ala-Mukhanov-potential}).

\subsection{Early power law inflation}

We are going to study spatially flat FRW cosmological models
governed by the system of equations
\begin{equation}
\frac{\dot{a}^{2}}{a^{2}}=\frac{1}{3M_{p}^{2}}\rho
\label{cosm-phi}
\end{equation}
and (\ref{rho1})-(\ref{phi1}), and in this subsection we will
consider the fine tuned case $\delta =0$. As
$M^{4}e^{-2\alpha\phi/M_{p}}\gg Max\left(|V_{1}|,
|V_{2}|/b_{g}\right)$, the effective potential
(\ref{eff-L-ala-Mukhanov-potential}) behaves as the exponential
potential $V_{eff}^{(0)}\approx
\frac{1}{4b_{g}}M^{4}e^{-2\alpha\phi/M_{p}}$. So, as $\phi\ll
-M_{p}$  the model describes the well studied power law inflation
of the early universe\cite{power-law} if $0<\alpha < 1/\sqrt{2}$ :
\begin{equation}
a(t)=a_{in}\left(\frac{t}{t_{in}}\right)^{1/2\alpha^{2}}, \qquad
\phi(t)
=\frac{M_p}{\alpha}ln\left(\frac{\alpha^2M^2t}{M_p\sqrt{b_g(3-2\alpha^2)}}\right)
. \label{p-l-solution}
\end{equation}
In the phase plane $(\phi,\dot{\phi})$ there is only one phase
curve representing these solutions and it plays the role of the
attractor\cite{Halliwell} for all other solutions with arbitrary
initial values of $\phi_{in}$ and $\dot{\phi}_{in}$. Excluding
time from $\phi(t)$ and $\dot{\phi}(t)$ we obtain the equation of
the attractor in the phase plane:
\begin{equation}
\frac{\sqrt{b_g}}{M^2}\,\dot{\phi}=\frac{\alpha}{\sqrt{3-2\alpha^2}}e^{-\alpha\phi/M_p}.
\label{p-l-phase-plane}
\end{equation}
 With our choice of $\alpha =0.2$ the
equation-of-state for the attractor is $w\approx -0.95$.

  Solutions corresponding to different independent initial values
$\phi_{in}$ and $\dot{\phi}_{in}$ have been studied numerically in
Ref.\cite{GK9}: shapes of the phase curves are characterized by
much steeper (almost vertical) approaching the attractor than the
exponential shape of the decay of the attractor itself. Graceful
exit from the inflationary regime was demonstrated in detail in
Ref.\cite{GK9} as well.

\subsection{Quintessential inflation type of scenario}

The model (\ref{totaction}) provides a possibility of a
quintessential inflation type of scenario\cite{Quint-infl} where
the early power law inflation ends with a graceful exit to a
quintessence-like epoch. This was demonstrated in Ref.(\cite{GK9})
in detail. Here we shortly illustrate this possibility in the case
$V_1>0$ and $b_{g}V_{1}\geq 2V_{2}$.

In this model
 the effective potential (\ref{eff-L-ala-Mukhanov-potential}) is a
monotonically decreasing function of $\phi$.   As $\phi\ll -M_{p}$
the model describes the power law inflation, similar to what we
discussed in the  previous subsection. Applying this model to the
cosmology of the late time universe and assuming that the scalar
field $\phi\rightarrow\infty$ as $t\rightarrow\infty$, it is
convenient to represent the effective potential
(\ref{eff-L-ala-Mukhanov-potential}) in the form
\begin{equation}
V_{eff}^{(0)}(\phi)=\Lambda +V_{q-l}(\phi)
\label{rho-without-ferm}
\end{equation}
with the definition
\begin{equation}
 \Lambda
=\frac{V_{1}^{2}} {4(b_{g}V_{1}-V_{2})}. \label{lambda}
\end{equation}
Here $\Lambda$ is the positive cosmological constant (see
(\ref{bV1>V2})) and
\begin{equation}
V_{q-l}(\phi)
=\frac{(b_{g}V_{1}-2V_{2})V_{1}M^{4}e^{-2\alpha\phi/M_{p}}+
(b_{g}V_{1}-V_{2})M^{8}e^{-4\alpha\phi/M_{p}}}
{4(b_{g}V_{1}-V_{2})[b_{g}(V_{1}+
M^{4}e^{-2\alpha\phi/M_{p}})-V_{2}]},
\label{V-quint-without-ferm-delta=0}
\end{equation}
that is the evolution of the late time universe  is governed both
by the cosmological constant $\Lambda$ and by the
quintessence-like potential $V_{q-l}(\phi)$.

Thus  the effective potential (\ref{eff-L-ala-Mukhanov-potential})
provides a possibility for a toy cosmological scenario which
passes through 3 stages: an early power law inflation,
quintessence and ends with a cosmological constant $\Lambda$.
Recall that the $\phi$-dependence of the effective potential
(\ref{eff-L-ala-Mukhanov-potential}) appears here only as the
result  of the spontaneous breakdown of the global scale symmetry
(\ref{st}) which in the Einstein frame is just a shift
symmetry(\ref{phiconst}).

\subsection{A tiny CC without fine tuning of dimensionfull parameters}

 In the scalar field models of dark energy, an
interesting feature of TMT consists in a possibility to provide a
small value of the cosmological constant without fine tuning of
dimensionfull parameters. If in the model of the previous
subsection the parameter $V_2<0$ and $|V_2|\gg b_gV_1$ then
$\Lambda$ can be very small without the need for $V_1$ and $V_2$
to be very small. For example, if $V_{1}$ is determined by the
 energy scale of electroweak symmetry breaking $V_{1}\sim
(10^{3}GeV)^{4}$ and $V_{2}$ is determined by the Planck scale
$V_{2} \sim (10^{18}GeV)^{4}$ then $\Lambda\sim (10^{-3}eV)^{4}$.
Along with such a seesaw  mechanism\cite{G1}, \cite{seesaw}, there
exists another way to explain the smallness of $\Lambda$. As one
can see from Eq.(\ref{lambda}),  the value of $\Lambda$ appears to
be inverse proportional to the dimensionless parameter $b_g$ which
characterizes the relative strength of the 'manifold' and
'metrical' parts of the gravitational action (see $S_g$ in
(\ref{totaction})). If for example $V_{1}\sim (10^{3}GeV)^{4}$
then for getting $\Lambda_1\sim (10^{-3}eV)^{4}$ one should assume
that $b_{g}\sim 10^{60}$. Such a large value of $b_{g}$ (see
Eq.(\ref{L1L2})) permits to formulate  {\em a correspondence
principle}\cite{GK9} {\em between TMT and regular (i.e.
one-measure) field theories}: when $\zeta/b_g\ll 1$ then one can
neglect the gravitational term in $L_1$ with respect to that in
$L_2$ (see Eq.(\ref{L1L2}) or Eq.(\ref{S-model-scalar.f.}) or
Eq.(\ref{totaction})). More detailed
analysis\cite{GK9},\cite{GK11} shows that in such a case the
manifold volume measure $\Phi =\zeta \sqrt{-g}$ has no a dynamical
effect and TMT is reduced to GR.

\section{Early TMT cosmology with no fine tuning: absence of the
initial singularity of the curvature}

 \subsection{General analysis and numerical solutions}

Let us now return to the general case of the model of Sec.5 with
no fine tuning of the parameters $b_g$ and $b_{\phi}$, i.e. the
parameter $\delta$, defined by Eq.(\ref{delta}), is now non zero.
The dynamics of the spatially flat FRW cosmology is described by
Eqs.(\ref{rho1})-(\ref{phi1}) and (\ref{cosm-phi}). Let us start
from the analysis of Eq.(\ref{phi1}). The interesting feature of
this equation is that
 each of the factors $Q_{i}(\phi,X)$ \, ($i=1,2,3$) can get
 to zero and this effect depends on the range of the parameter
 space chosen. This is the
origin of drastic {\it changes of the topology of the phase plane}
comparing with the fine tuned model of Sec.5.2.

\begin{figure}[htb]
\begin{center}
\includegraphics[width=12.0cm,height=9.0cm]{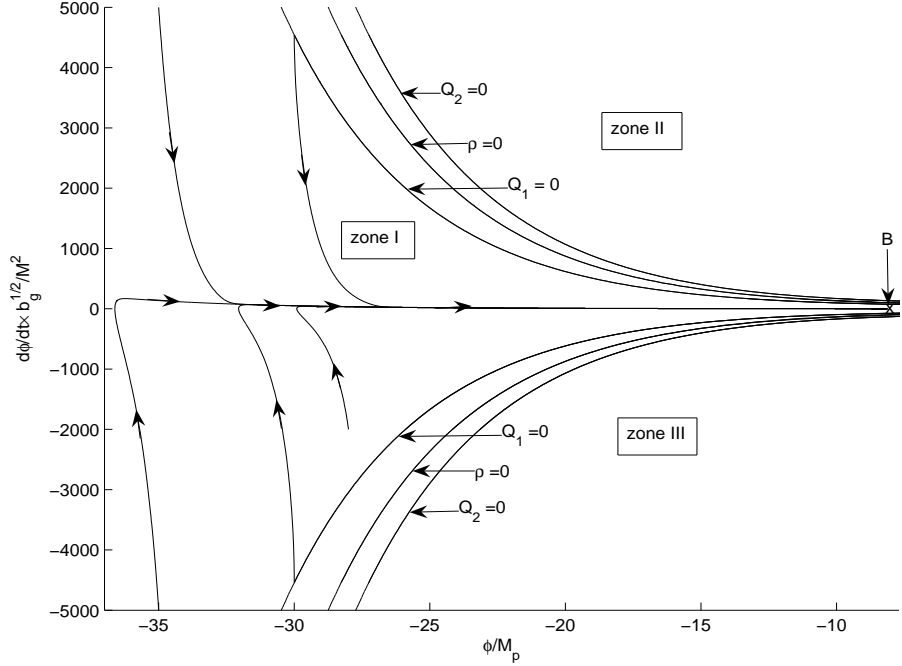}
\end{center}
\caption{The phase portrait for the model with $\delta =0.1$,
$\alpha =0.2$, $V_{1}=-30M^{4}$ and $V_{2}=-50b_{g}M^{4}$.  All
phase curves demonstrate the attractor behavior similar to that in
the fine tuned case  $\delta =0$, Fig.2. One can see that the
attractor does not intersect the line $Q_1=0$, see also Fig.10.
The location of the oscillatory regime marked by point $B$ is
exactly the same as in the model with $\delta =0$. The essential
difference consists in a novel topological structure: in the
neighborhood of the line $Q_1=0$ all the phase curves exhibit a
repulsive behavior from this line and therefore points of the line
$Q_1=0$ are dynamically unachievable. Hence the phase curves
cannot be continued infinitely to the past.}\label{fig3}
\end{figure}

\begin{figure}[htb]
\begin{center}
\includegraphics[width=12.0cm,height=5.0cm]{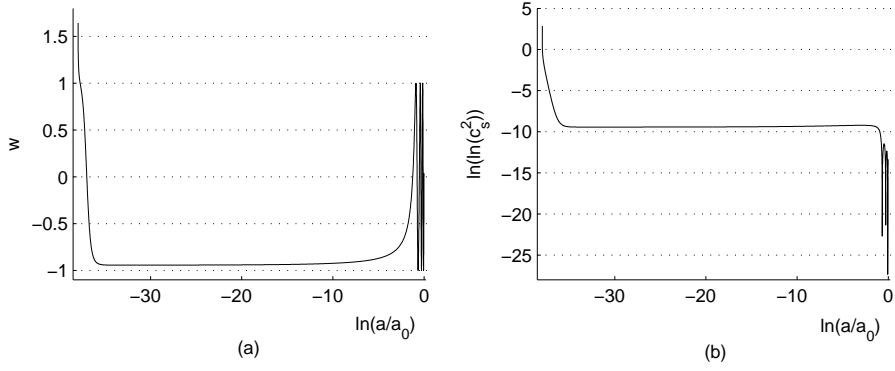}
\end{center}
\caption{Typical scale factor dependence of the equation-of-state
$w$ and the squared sound speed $c^2_s$ for the phase curves
starting from  points very close to the line $Q_1=0$ (in this
figure - for $\phi_{in}=-30M_p$ and $\dot{\phi}_{in}=
4540.7569b_g^{-1/2}M^2$). The squared sound speed in the starting
point is $c^2_s\approx exp(exp (2.86))\approx 3.8\cdot 10^7$. It
is not a problem to obtain more than 75 e-folds during the power
law inflation (the region of $w\approx -0.95$ in Fig.(a)) just by
choosing a larger absolute value $|\phi_{in}|$. We have chosen
$\phi_{in}=-30M_p$ because this allows to show more details in
these and subsequent graphs. }\label{fig4}
\end{figure}

\begin{figure}[htb]
\begin{center}
\includegraphics[width=12.0cm,height=5.0cm]{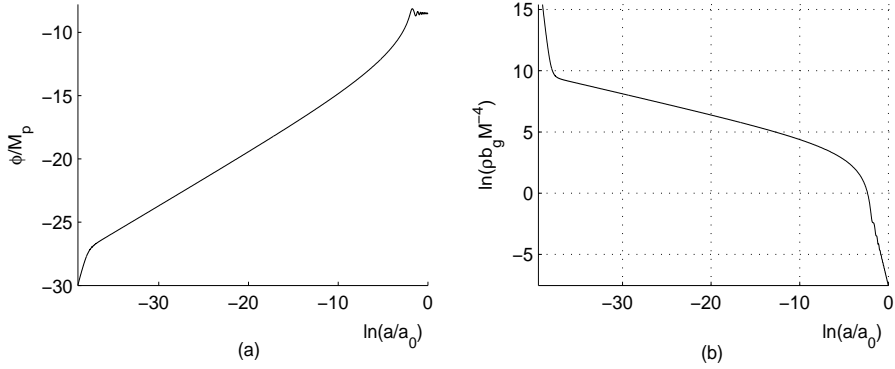}
\end{center}
\caption{Typical scale factor dependence of $\phi$ and $\rho$ for
the phase curves starting from  points very close to the line
$Q_1=0$ (in this figure - for the same point as in
Fig.4.}\label{fig5}
\end{figure}

For $Q_{1}\neq 0$, Eqs. (\ref{phi1}), (\ref{cosm-phi}) result in
the well known equation\cite{Vikman}
\begin{equation}
\ddot{\phi}+
\frac{\sqrt{3\rho}}{M_p}c_s^2\dot{\phi}+\frac{\rho_{,\phi}}{\rho_{,X}}
=0, \label{phantom-phi}
\end{equation}
where $c_s$ is the effective sound speed of perturbations\cite{GM}
\begin{equation}
c_s^2=\frac{p_{,X}}{\rho_{,X}}=\frac{Q_2}{Q_1}, \label{c-s-2}
\end{equation}
\begin{equation}
\frac{\rho_{,\phi}}{\rho_{,X}}=- \frac{\alpha}{M_{p}}\cdot
\frac{Q_3}{Q_1}\cdot M^{4}e^{-2\alpha\phi/M_{p}}; \label{Q-rho-p}
\end{equation}
$\rho$ and $p$ are defined by Eqs.(\ref{rho1}), (\ref{p1}) and
$Q_i$ $(i=1,2,3)$ - by Eqs.(\ref{Q1})-(\ref{Q3}).

It follows from the definitions of $Q_1$ and $Q_2$ that $c_s^2>1$
when $Q_1>0$ and $X>0$ (that implies $Q_2>0$). Therefore in the
cosmological FRW background,  the sound speed of perturbations can
be bigger than speed of light\cite{Babichev-Mukh}.

In the model with $V_1<0$ and $V_2<0$,  the structure of the phase
plane is presented in Fig.3. With a choice of the parameters made,
the following condition is satisfied
\begin{equation}
(b_g+b_{\phi})V_1-2V_2>0 \label{cond-for-phase-structure}.
\end{equation}
As we will see in Sec.7, the latter is the characteristic
condition which determines the structure of the phase plane.
 By the two branches
of the line $Q_1=0$, the phase plane is divided into three large
dynamically disconnected zones. In the most part of two of them
(II and III), the energy density is negative ($\rho <0$). In the
couple of regions between the lines $Q_1=0$ and $\rho =0$ that we
refer for short as ($Q_1\rightarrow\rho$)-regions, $\rho
>0$ but $Q_1<0$. Typical scale factor dependence of the
equation-of-state $w$, the sound speed of perturbations $c_s^2$,
the inflaton $\phi$ and the energy density $\rho$ are presented in
Figs.4  and 5. It is not a problem to obtain more than 75 e-folds
during the power law inflation (the region of $w\approx -0.95$ in
Fig.4a) just by choosing a larger absolute value $|\phi_{in}|$. We
have chosen $\phi_{in}=-30M_p$ because this allows to show more
details in these graphs.

The zone I of the phase plane (where $Q_1>0$, $\rho
>0$ and $Q_2>0$) is of a great cosmological interest:



a) \, Similar to what we have seen in the fine tuned case in
 Sec.5.2, all the phase
 curves start with very steep approach to an attractor.  The nonlinearity in $X$
 does not allow to obtain an exact analytic solution in the model under
consideration and therefore we have no here the analytic equation
of the attractor. But one can show\cite{GK9} that, with our choice
of the parameters $\alpha$ and $\delta$, the emergence of the
additional terms $\propto\delta X$ and $\propto\delta^2X^2$ in the
model under consideration results in small enough corrections to
the equation of the attractor in comparison with
Eq.(\ref{p-l-phase-plane}) of the fined tuned model of Sec.IVB.
For our qualitative analysis below one can use
Eq.(\ref{p-l-phase-plane}) as a good approximation to the true
attractor equation.


b) \, Comparing the  equation of the line $Q_1=0$, which for
$\phi\ll-M_p$ can be written in the form
\begin{equation}
\frac{\sqrt{b_g}}{M^2}\,\dot{\phi}=\pm\frac{1}{\delta}
\sqrt{\frac{2}{3}(2-\delta)}\cdot e^{-\alpha\phi/M_p}+  O
\left(\frac{V_1}{M^4}e^{\alpha\phi/M_p} \right),
 \label{Q1=0}
\end{equation}
with the equation of the attractor which approximately coincides
with Eq.(\ref{p-l-phase-plane}), we see that the upper brunch of
 the line $Q_1=0$ has actually the same form of a decaying
 exponent as the attractor, but the factor in front of the exponent
  in Eq.(\ref{Q1=0})
 is about $10^2$ times bigger than in Eq.(\ref{p-l-phase-plane}).
 The above analytic estimations are confirmed by the numerical solutions
 as one can see in Fig.3.
 This means that the attractor does not intersect the line
 $Q_1=0$. Therefore all the phase curves starting in zone I arrive
 at
 the attractor (of course asymptotically).


c) \, In the neighborhood of the line $Q_1=0$ all the phase curves
exhibit  a {\it repulsive} behavior from this line. In other
words, the shape of two branches of $Q_1=0$ do not allow a
classical dynamical continuation of the phase curves backward in
time without crossing the classical barrier formed by the line
$Q_1=0$. This is true for all finite values of the initial
conditions $\phi_{in}$, $\dot{\phi}_{in}$ in zone I.


d) \,As one can see from  Fig.4a, the initial stage of evolution
is very much different from the subsequent one, that is a power
law inflation. This fact may have
 a relation to the results of the study of
completeness of inflationary cosmological models in past
directions\cite{BGV}.


e) \,If the phase curves start from points
$(\phi_{in},\dot{\phi}_{in})$ in zone I very close to the line
$Q_1=0$ then the sound speed of perturbations has huge values at
the beginning of the evolution, see Figs.4b. However in the power
law inflation stage, $c_s$ is too close to the speed of light and
appears to be unable to increase the tensor-to-scalar perturbation
ratio\cite{GM}.


In two regions between the lines $Q_1=0$ and $Q_2=0$ that we refer
for short as ($Q_1\rightarrow Q_2$)-regions,, the squared sound
speed of perturbations is negative, $c_s^2<0$. This means that on
the right hand side of the classical barrier $Q_1=0$, the model is
absolutely unstable. Moreover, this pure imaginary sound speed
becomes infinite in the limit $Q_1\rightarrow 0-$. Thus the
branches of the line $Q_1=0$ divide zone I (of the classical
dynamics) from the ($Q_1\rightarrow Q_2$)-regions where the
physical significance of the model is unclear. Note that the line
$\rho =0$ divides the ($Q_1\rightarrow Q_2$)-regions into two
subregions with opposite signs of the classical energy density.

Thus the structure of the phase plane yields a conclusion that
 the starting point of the classical history in the phase
plane can be only in zone I and {\bf the line $Q_1 =0$ is the
limiting set of points where the classical history might begin}.
For any finite initial values of $\phi_{in}$ and $\dot{\phi}_{in}$
at the initial cosmic time $t_{in}$, the  duration $t_{in}-t_{s}$
of the continuation of the evolution into the past up to the
moment $t_{s}$ when the phase trajectory arrives the line $Q_1
=0$, is finite.

\subsection{Analysis of the initial singularity}

Let us analyze what happens  as $t\to t_{s}$ (and
$Q_1(\phi,\dot{\phi})\to 0$) if this continuation to the past
starts from a point in zone I of the phase plane with finite
initial values $\phi_{in}$ and $\dot{\phi}_{in}$ . First note that
the energy density $\rho_{s}=\rho(t_{s})$ and the pressure
$p_{s}=p(t_{s})$ are finite in all the points of the line $Q_1 =0$
with finite coordinates $\phi_s=\phi(t_s)$,
$\dot{\phi_s}=\dot{\phi}(t_s)$, that it is easy to see from
Eqs.(\ref{rho1}), (\ref{p1}) and (\ref{Q1}). The strong energy
condition is satisfied in regions of zone I close to the line $Q_1
=0$ including the line itself. In fact, for any unit time-like
vector $t^{\mu}$ we have on the line $Q_1 =0$
\begin{equation}
\left(T_{\mu\nu}^{eff}-\frac{1}{2}\tilde{g}_{\mu\nu}T^{eff}\right)t^{\mu}t^{\nu}=\frac{1}{2}(\rho_s
+3p_s)\approx
\frac{2b_{\phi}}{(b_g-b_{\phi})^2}M^{4}e^{-2\alpha\phi_s/M_{p}}>0,
\label{energy-codition}
\end{equation}
where $T_{\mu\nu}^{eff}$ is defined in Eq.(\ref{Tmnfluid}),
$T^{eff}\equiv\tilde{g}^{\mu\nu}T_{\mu\nu}^{eff}$ and we have
taken into account our choice of the parameters ($b_g>0$,
$b_{\phi}>0$ and Eq.(\ref{bV1>V2})) and assumed that
$M^{4}e^{-2\alpha\phi_s/M_{p}}\gg |V_1|$. This result is in the
total agreement with the numerical solutions of the previous
subsection. In particular, the described analytic approximation on
the line $Q_1 =0$ yields $w_s=p_s/\rho_s\approx 5/3$ which is in a
very good agreement with the numerical results obtained in regions
of zone I close enough to the line $Q_1 =0$, see Fig.4a.

It follows from the Einstein equations (\ref{cosm-phi}) and
\begin{equation}
\frac{\ddot{a}}{a}=-\frac{1}{6M_p^2}(\rho
+3p)\label{ddotEinsteinEq}
\end{equation}
that the first and second time derivatives of the scale factor,
$\dot{a}_{s}=\dot{a}(t_{s})$ and $\ddot{a}_{s}=\ddot{a}(t_{s})$,
and therefore the curvature, are finite on the line $Q_1 =0$. The
time derivative of the energy density also approaches a finite
value
\begin{equation}
\dot{\rho}_s=3\frac{\dot{a}_{s}}{a_s}(\rho_s +p_s)<\infty
\label{dot-rho-finite}
\end{equation}
but $\dot{p}\to -\infty$ \, as \, $t\to t_s $ because
$|\ddot{\phi}|\to\infty$.

The detailed analysis shows that
\begin{equation}
a(t)\approx a_s +A(t-t_s)^{5/2} \quad \text{and}\quad
\dot{\phi}=\dot{\phi}_s\pm \frac{M_p}{2\alpha}\sqrt{2B(t-t_s)}
\quad \text{as} \quad t\to t_s, \label{a(t)-n}
\end{equation}
where $A>0$ and $B>0$i are  constants in time.

Therefore although the scalar curvature
\begin{equation}
R=-6(\frac{\ddot{a}}{a}+\frac{\dot{a}^2}{a^2})\label{ScalarCurvature}
\end{equation}
is finite as $t\to t_s$ but its time derivative is singular:
\begin{equation}
\dot{R}\approx -6\frac{\dddot{a}}{a}\to -\infty \qquad \text{as}
\qquad t\to t_s\label{ScalarCurvature}
\end{equation}

 This type of singularity we discover here in the framework of
the dynamical model is present in the classification of "sudden"
singularities given by Barrow\cite{Barrow1}.

Finally we want to discuss possible scales of the energy when the
initial conditions are close to the line $Q_1 =0$. If
$\phi_{in}\approx\phi_s\ll -M_p$ then approximately
\begin{equation}
\rho_{in}\approx\rho_s\approx\frac{(b_g+b_{\phi})^2}{12b_g(b_g-b_{\phi})^2}M^4e^{-2\alpha\phi_s/M_p}
\label{rho-very-big}
\end{equation}
Therefore depending on the parameters and initial conditions, the
described mild singular initial behavior is possible for the
energy densities close to the Planck scale as well as for the
energy densities much lower the quantum Planck era. It is
interesting that the singularity of the time derivative of the
curvature on the line $Q_1 =0$ is accompanied with singularity of
the sound speed of perturbations $c_s^2
=Q_2/Q_1\sim\ddot{\phi}\sim(t-t_s)^{-1/2}$. Therefore generation
of any mode of scalar fluctuations in states extremely close to
the line $Q_1=0$ requires extremely large energy. Thus the initial
state formed in the close neighborhood of the line $Q_1=0$ must be
practically the ground state. This allows to hope that the effect
$c_s^2\to\infty$ could help to solve the problem of the initial
conditions in inflationary cosmology.

 \begin{figure}[htb]
\begin{center}
\includegraphics[width=13.0cm,height=8.0cm]{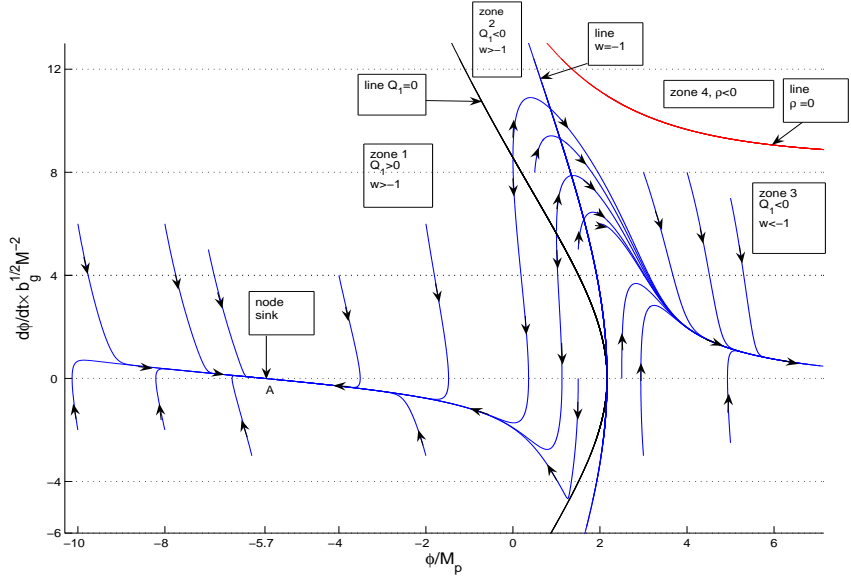}
\end{center}
\caption{The phase portrait (in the phase plane
($\phi$,$\dot{\phi})$) for the model with $\alpha =0.2$, $\delta
=0.1$, $V_{1}=10M^{4}$ and $V_{2}=9.9b_{g}M^{4}$. The region with
$\rho >0$ is divided into two dynamically disconnected regions by
the line $Q_{1}(\phi,\dot{\phi})=0$. To the left of this line
$Q_{1}>0$ (the appropriate zone we call zone 1) and to the right
\, $Q_{1}<0$.  The $\rho
>0$ region to the right of the line $Q_{1}(\phi,\dot{\phi})=0$ is
divided into two zones (zone 2 and zone 3) by the line $Q_2=0$
(the latter coincides with the line where $w=-1$). In zone 2 \,
$w>-1$ but $c_s^2<0$. In zone 3 \,$w<-1$ and $c_s^2>0$. Phase
curves started in zone 2 cross the line $w=-1$. All phase curves
in zone 3 exhibit processes with super-accelerating expansion of
the universe. Besides all the phase curves in zone 3 demonstrate
 attractor behavior to the line which asymptotically, as
$\phi\to\infty$, approaches  the straight line
$\dot{\phi}=0$.}\label{fig6}
\end{figure}
\begin{figure}[htb]
\includegraphics[width=10.0cm,height=6.0cm]{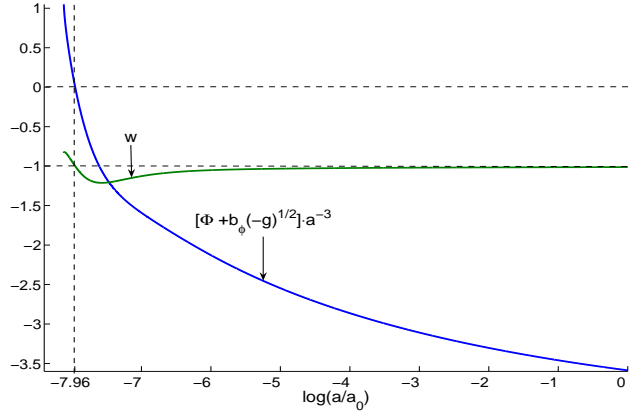}
\caption{For the same model as in Fig. 6 and with the initial
conditions $\phi_{in}=M_{p}$, $\dot\phi_{in} =5.7M^2/\sqrt{b_g}$:
crossing the phantom divide $w=-1$ and changing sign of the total
volume measure $(\Phi +b_{\phi}\sqrt{-g})$ in the scalar field
$\phi$ kinetic term (in the underlying action (\ref{totaction}))
occur simultaneously.}\label{fig7}
\end{figure}

\section{Sign indefiniteness of the manifold volume measure
 as the origin of a phantom dark energy}

 We turn now to the non fine-tuned case of the model of Sec.VII
applied to the spatially flat universe. We start from a short
review of our recent results\cite{GK9} concerning  qualitative
structure of the appropriate dynamical system which consists of
Eqs. (\ref{phi1}), (\ref{cosm-phi}) where the energy density
$\rho$ is defined by Eq.(\ref{rho1}). The case of the interest of
this section is realized when the parameters of the model satisfy
the condition
\begin{equation}
(b_g+b_{\phi})V_1-2V_2<0 \label{cond-for-phase-structure-1}
\end{equation}
In this case the phase plane has a very interesting structure
presented in Fig.6. Recall that the functions $Q_1$, $Q_2$, $Q_3$
are defined by Eqs.(\ref{Q1})-(\ref{Q3}).

We are interested in the equation of state $w=p/\rho<-1$, where
 pressure $p$ and energy density $\rho$ are given by
Eqs.(\ref{rho1}) and (\ref{p1}). The line indicated in Fig. 6 as
"line $w=-1$" coincides with the line $Q_2(\phi,X)=0$ because
\begin{equation}
w+1=\frac{X}{\rho}\cdot\frac{Q_2}{\left[b_g\left(M^4e^{-2\alpha\phi/M_p}+V_1\right)-V_2\right]}
\label{w+1}
\end{equation}
 Phase curves in zone 3 correspond to the
cosmological solutions  with the equation of state $w<-1$. In zone
2, $w>-1$ but this zone has no physical meaning since the squared
sound speed of perturbations
\begin{equation}
c_s^2=\frac{Q_2}{Q_1} \label{sound}
\end{equation}
is negative in zone 3. But in zone 2, $c_s^2>0$. Some details of
 numerical solutions describing the cross of the phantom divide
$w=-1$ are presented in Fig. 7.

Note that the superaccelerating cosmological expansion is obtained
here without introducing an explicit phantom scalar field into the
underlying action (\ref{totaction}). In Ref.\cite{GK9} we have
discussed this effect from the point of view of the effective
k-essence model realized in the Einstein frame when starting from
the action (\ref{totaction}). A deeper analysis\cite{GK11} of the
same effect yields the conclusion that the true and profound {\em
origin of the appearance of an effective phantom dynamics in our
model is sign-indefiniteness of the manifold volume measure
$\Phi$}. In fact, using the constraint (\ref{constraint2}),
Eqs.(\ref{w+1}) and (\ref{ct}) it is easy to show that
\begin{equation}
\Phi
+b_{\phi}\sqrt{-g}=(w+1)\,\frac{\rho}{4X}\,\frac{[M^4e^{-2\alpha\phi/M_p}+V_1+\delta\cdot
b_gX]}{[b_g(M^4e^{-2\alpha\phi/M_p}+V_1)-V_2]}\, a^3,
\label{kinetic-measure}
\end{equation}
where $a$ is the scale factor. The expression in the l.h.s of this
equation is the total volume measure of the $\phi$ kinetic term in
the underlying action (\ref{totaction}):
\begin{equation}
\int d^{4}x (\Phi
+b_{\phi}\sqrt{-g})\frac{1}{2}g^{\mu\nu}\phi_{,\mu}\phi_{,\nu}
\label{kinetic-total-measure}
\end{equation}
 The sign of this volume measure
coincides with the sign of $w+1$ as well as with the sign of the
function $Q_2$ (see Eq.(\ref{w+1})). In Fig. 8 we present the
result of numerical solution for the scale factor dependence of
$w$ and $(\Phi +b_{\phi}\sqrt{-g})/a^3$.  Thus {\em crossing the
phantom divide occurs when  the total volume measure of the $\phi$
kinetic term in the underlying action changes sign from positive
to negative for dynamical reasons}. This dynamical effect appears
here as a dynamically well-founded alternative to the usually
postulated phantom kinetic term of a scalar field
Lagrangian\cite{Phantom-usual}.

\section{Including fermions into the model II}
\subsection{Description of the model}

The scale invariant TMT model of Sec.5 allows a generalization by
involving fermions as well as gauge and Higgs fields in such a way
that the standard particle model can be obtained. In this short
review, to simplify the presentation of the main results we will
study an Abelian gauge model which does not include the Higgs
fields and quarks and chiral properties  of fermions are ignored.
In this simplified model the mass terms for Dirac spinors are
included by hand from the very beginning into the Lagrangians
$L_{1}$ and $L_{2}$.  The matter content of our model includes the
dilaton scalar field $\phi$, two so-called primordial fermion
fields (the primordial neutrino $N$ and the  primordial electron
$E$) and electromagnetic field $A_{\mu}$. The latter is included
in order to display the reasons why the gauge fields dynamics is
canonical. Generalization to non-Abelian gauge models including
also quarks, Higgs fields, the Higgs mechanism of mass generation
and taking into account chiral properties  of fermions is
straightforward, see Ref.\cite{GK5}.

We allow in both $L_{1}$ and $L_{2}$ all the usual terms
considered in standard field theory models in curved space-time.
It is convenient to represent the action in the following form:
\begin{eqnarray}
S=S_g+S_{\phi}+S_f+S_{em} \quad \text{where}\label{totaction-1}\\
S_f= \int d^{4}x e^{\alpha\phi /M_{p}}(\Phi +k\sqrt{-g})
\frac{i}{2}\sum_{i}\overline{\Psi}_{i}
\left(\gamma^{a}e_{a}^{\mu}\overrightarrow{\nabla}^{(i)}_{\mu}-
\overleftarrow{\nabla}^{(i)}_{\mu}\gamma^{a}e_{a}^{\mu}\right)\Psi_{i}
\label{Sf}\\
     -\int d^{4}xe^{\frac{3}{2}\alpha\phi /M_{p}}
\left[(\Phi +h_{E}\sqrt{-g})\mu_{E}\overline{E}E +(\Phi
+h_{N}\sqrt{-g})\mu_{N}\overline{N}N \right] \nonumber\\
S_{em}=-\frac{1}{4}\int
d^{4}x\sqrt{-g}g^{\alpha\beta}g^{\mu\nu}F_{\alpha\mu}F_{\beta\nu}
\label{Sem}
\end{eqnarray}
and $S_g$, $S_{\phi}$ are determined in Eq.(\ref{totaction}) where
we take for simplicity $b_g=b_{\phi}=b$. \, $\Psi_{i}$ ($i=N,E$)
is the general notation for the primordial fermion fields $N$ and
$E$;
$F_{\alpha\beta}=\partial_{\alpha}A_{\beta}-\partial_{\beta}A_{\alpha}$;
$\mu_{N}$ and $\mu_{E}$ are the mass parameters;
$\overrightarrow{\nabla}^{(N)}_{\mu}=\vec{\partial}+
\frac{1}{2}\omega_{\mu}^{cd}\sigma_{cd}$,
$\overrightarrow{\nabla}^{(E)}_{\mu}=\vec{\partial}+
\frac{1}{2}\omega_{\mu}^{cd}\sigma_{cd}+ieA_{\mu}$;
 $R(\omega ,V)
=e^{a\mu}e^{b\nu}R_{\mu\nu ab}(\omega)$ is the scalar curvature;
$e_{a}^{\mu}$ and $\omega_{\mu}^{ab}$ are the vierbein and
spin-connection; $g^{\mu\nu}=e^{\mu}_{a}e^{\nu}_{b}\eta^{ab}$ and
$R_{\mu\nu ab}(\omega)=\partial _{mu}\omega_{\nu
ab}+\omega^{c}_{\mu a}\omega_{\nu cb}-(\mu \leftrightarrow\nu)$.
If Higgs field were included into the model then $V_{1}$ and $V_2$
would turn into functions of the Higgs field. Constants $b, k,
h_{N}, h_{E}$  are non specified dimensionless real parameters and
we will only assume that they
 have close orders of magnitude
\begin{equation}
b\sim k\sim h_{N}\sim h_{E}. \label{sim-parameters}
\end{equation}

The action (\ref{totaction-1}) is evidently gauge invariant and
besides it is invariant under the global scale transformations:
\begin{eqnarray}
    &&e_{\mu}^{a}\rightarrow e^{\theta /2}e_{\mu}^{a}, \quad
\omega^{\mu}_{ab}\rightarrow \omega^{\mu}_{ab}, \quad
\varphi_{a}\rightarrow \lambda_{ab}\varphi_{b}\quad
A_{\alpha}\rightarrow A_{\alpha}, \nonumber
\\
 &&\phi\rightarrow
\phi-\frac{M_{p}}{\alpha}\theta ,\quad \Psi_{i}\rightarrow
e^{-\theta /4}\Psi_{i}, \quad \overline{\Psi}_{i}\rightarrow
e^{-\theta /4} \overline{\Psi}_{i}, \label{stferm}
\\
 &&\text{where} \quad \theta =\text{const}, \quad \lambda_{ab}=\text{const} \quad and
 \quad \det(\lambda_{ab})=e^{2\theta},
\nonumber
\end{eqnarray}
 We have chosen the kinetic term of $A_{\mu}$ in the conformal
invariant form which is possible  if it is coupled only to the
measure $\sqrt{-g}$. Introducing the coupling of this term to the
measure $\Phi$ would lead to the nonlinear field strength
dependence  in the $A_{\mu}$ equation of motion. One of the
possible consequences of this  may be  non positivity of the
energy density. Another consequence is a possibility of certain
unorthodox effects, like space-time variations of the effective
fine structure constant. This subject deserves a special study but
it is out of the purposes of this paper where the model is studied
setting $F_{\mu\nu}\equiv 0$ in the solutions.

Variation of the measure fields $\varphi_{a}$  yields
Eq.(\ref{varphi}) where $L_{1}$ is defined as the part of the
integrand of the action (\ref{totaction-1})-(\ref{Sem}) coupled to
the measure $\Phi$. The appearance of the integration constant
$sM^{4}$ in Eq.(\ref{varphi}) spontaneously breaks the global
scale invariance (\ref{stferm}). In what follows we choose $s=+1$.

Except for the $A_{\mu}$ equation, all other equations of motion
resulting from (\ref{totaction-1})-(\ref{Sem}) in the first order
formalism contain terms proportional to $\partial_{\mu}\zeta$ that
makes the space-time non-Riemannian and equations of motion - non
canonical. However, with the new set of variables ($\phi$ and
$A_{\mu}$ remain unchanged)
\begin{eqnarray}
&&\tilde{e}_{a\mu}=e^{\frac{1}{2}\alpha\phi/M_{p}}(\zeta
+b)^{1/2}e_{a\mu}, \quad
\tilde{g}_{\mu\nu}=e^{\alpha\phi/M_{p}}(\zeta +b)g_{\mu\nu},
\nonumber\\
&&\Psi^{\prime}_{i}=e^{-\frac{1}{4}\alpha\phi/M_{p}} \frac{(\zeta
+k)^{1/2}}{(\zeta +b)^{3/4}}\Psi_{i} , \quad i=N,E \label{ctferm}
\end{eqnarray}
which we call the Einstein frame,
 the spin-connections become those of the
Einstein-Cartan space-time. Since $\tilde{e}_{a\mu}$,
$\tilde{g}_{\mu\nu}$, $N^{\prime}$ and $E^{\prime}$ are invariant
under the scale transformations (\ref{stferm}), spontaneous
breaking of the scale symmetry  is reduced in the new variables to
the  spontaneous breaking of the shift symmetry (\ref{phiconst}).

After the change of variables (\ref{ctferm}) to the Einstein frame
and some simple algebra, the gravitational equations  take the
standard GR form (\ref{gef}) where the energy-momentum tensor
$T_{\mu\nu}^{eff}$ is now
\begin{eqnarray}
T_{\mu\nu}^{eff}&=&\phi_{,\mu}\phi_{,\nu}-\frac{1}{2}
\tilde{g}_{\mu\nu}\tilde{g}^{\alpha\beta}\phi_{,\alpha}\phi_{,\beta}
+\tilde{g}_{\mu\nu}V_{eff}(\phi,\zeta;M)
\nonumber\\
&+&T_{\mu\nu}^{(em)}
+T_{\mu\nu}^{(ferm,can)}+T_{\mu\nu}^{(ferm,noncan)},
 \label{Tmn}
\end{eqnarray}
where $V_{eff}(\phi,\zeta;M)$ is defined by Eq.(\ref{Veff2}), \,
$T_{\mu\nu}^{(em)}$ is the canonical energy momentum tensor of the
electromagnetic field; $T_{\mu\nu}^{(ferm,can)}$ is the canonical
energy momentum tensor for (primordial) fermions $N^{\prime}$ and
$E^{\prime}$ in curved space-time (including also interaction of
$E^{\prime}$ with $A_{\mu}$). $T_{\mu\nu}^{(ferm,noncan)}$ is the
{\em noncanonical} contribution of the fermions into the energy
momentum tensor
\begin{equation}
 T_{\mu\nu}^{(ferm,noncan)}=-\tilde{g}_{\mu\nu}\Lambda_{dyn}^{(ferm)}
 \label{Tmn-noncan}
\end{equation}
where
\begin{equation}
\Lambda_{dyn}^{(ferm)}\equiv Z_{N}(\zeta)m_{N}(\zeta)
\overline{N^{\prime}}N^{\prime}+
Z_{E}(\zeta)m_{E}(\zeta)\overline{E^{\prime}}E^{\prime}
\label{Lambda-ferm}
\end{equation}
and the function $Z_{i}(\zeta)$ and fermion masses $m_{i}(\zeta)$
($i=N^{\prime},E^{\prime}$) are respectively
\begin{equation}
Z_{i}(\zeta)\equiv \frac{(\zeta -\zeta^{(i)}_{1})(\zeta
-\zeta^{(i)}_{2})}{2(\zeta +k)(\zeta +h_{i})}, \qquad
m_{i}(\zeta)= \frac{\mu_{i}(\zeta +h_{i})}{(\zeta +k)(\zeta
+b)^{1/2}} \label{Zeta&m}
\end{equation}
where
\begin{equation}
\zeta_{1,2}^{(i)}=\frac{1}{2}\left[k-3h_{i}\pm\sqrt{(k-3h_{i})^{2}+
8b(k-h_{i}) -4kh_{i}}\,\right].
 \label{zeta12}
\end{equation}

The noncanonical contribution $T_{\mu\nu}^{(ferm,noncan)}$ of the
fermions into the energy momentum tensor has the transformation
properties of a cosmological constant term but it is proportional
to fermion densities
$\overline{\Psi}^{\prime}_{i}\Psi^{\prime}_{i}$ \,
($i=N^{\prime},E^{\prime}$).

The "dilaton" $\phi$ field equation  in the Einstein frame reads
\begin{equation}
\Box\phi -\frac{\alpha}{M_{p}(\zeta +b)}
\left[M^{4}e^{-2\alpha\phi/M_{p}}-\frac{(\zeta -b)V_{1}(\upsilon)
+2V_{2}(\upsilon)}{\zeta +b}\right]= -\frac{\alpha
}{M_{p}}\Lambda_{dyn}^{(ferm)}, \label{phief+ferm1}
\end{equation}
where $\Box\phi =(-\tilde{g})^{-1/2}\partial_{\mu}
(\sqrt{-\tilde{g}}\tilde{g}^{\mu\nu}\partial_{\nu}\phi)$.

One can show that equations  for the primordial fermions in terms
of the variables (\ref{ctferm})
 take the standard form
of fermionic equations for $N^{\prime}$ and $E^{\prime}$ in the
Einstein-Cartan space-time  where the standard electromagnetic
interaction of $E^{\prime}$ is present too. Besides fermion
$4$-currents have the canonical form and are  covariantly
conserved. In what it concerns the fermion equations, all the
novelty
 as compared with the standard field theory
approach  consists of the form of the $\zeta$ depending "masses"
of the primordial fermions, second equation in Eq.(\ref{Zeta&m}).

The  electromagnetic field equations are canonical due to our
choice  for the appropriate term in the action (\ref{Sem}) to be
conformally invariant.

The  scalar field $\zeta\equiv\Phi/\sqrt{-g}$ is determined as a
function of the scalar field $\phi$ and
$\overline{\Psi}^{\prime}_{i}\Psi^{\prime}_{i}$ \,
($i=N^{\prime},E^{\prime}$) by the following constraint
\begin{equation}
\frac{1}{(\zeta
+b)^{2}}\left\{(b-\zeta)\left[M^{4}e^{-2\alpha\phi/M_{p}}+
V_{1}(\upsilon)\right]-2V_{2}(\upsilon)\right\}=
\Lambda_{dyn}^{(ferm)} \label{constraint}
\end{equation}
which is a straightforward generalization of the constraint
(\ref{constraint2}) in the presence of fermions (recall that we
have chosen here $\delta =0$).

One should point out an unexpected and very important fact, namely
that the same function $\Lambda_{dyn}^{(ferm)}$,
Eq.(\ref{Lambda-ferm}), emerges in
 three different places: \quad a) in the form of the noncanonical fermion contribution
 to the energy-momentum tensor, Eq.(\ref{Tmn-noncan}); \quad b)  in the
 effective Yukawa coupling of the dilaton $\phi$ to fermions
 (see the right hand side of Eq.(\ref{phief+ferm1})); \quad
c) as the right hand side of the constraint.

Note that the original action (\ref{totaction-1})-(\ref{Sem})
 contains exponents
of the scalar field $\phi$ and in particular the coupling of
fermions with $\phi$ is realized in
(\ref{totaction-1})-(\ref{Sem}) only through the exponents of
$\phi$. Nevertheless, except for the term
$M^{4}e^{-2\alpha\phi/M_{p}}$ originated by the scale symmetry
breaking, the equations of motion  in the Einstein frame do not
contain explicitly the exponents of $\phi$ . However, the
Yukawa-type coupling of the fermions to $\phi$ emerges in the
Einstein frame, see the r.h.s. of Eq.(\ref{phief+ferm1}). It is
interesting that non-explicit dependence on the exponent of $\phi$
in the equations of motion is actually present after solving the
constraint (\ref{constraint}) for $\zeta$. However this dependence
is again in the form of $M^{4}e^{-2\alpha\phi/M_{p}}$. Thus the
exponential $\phi$-dependence in the equations of motion results
only from the scale symmetry breaking. Recall that in the Einstein
frame the scale symmetry transformations (\ref{stferm}) are
reduced to the shift symmetry $\phi\rightarrow\phi + const$.

\subsection{Fermionic matter in normal conditions: reproducing GR
and fine tuning free decoupling from the quintessence field}

\subsubsection{Meaning of the constraint, regular fermions and reproducing Einstein equations}

The detailed analysis shows\cite{GK5},\cite{GK7} that  the l.h.s.
of the constraint (\ref{constraint}) has the order of magnitude
close to that of the (quintessence) dark energy density. At the
same time, in the presence of a single massive primordial fermion,
the r.h.s. of the constraint (\ref{constraint}) contains factor
$m_{i}(\zeta)\overline{\Psi}^{\prime}_{i}\Psi^{\prime}_{i}$, \,
 which have typical order of magnitude
of the fermion canonical energy density $T_{00}^{(ferm,can)}$. In
other words,  the constraint describes the local balance between
the fermion energy density and the scalar dark energy density in
the space-time region where the  wave function of the primordial
fermion is not equal to zero; this balance is realized due to the
factor $Z_{i}(\zeta)$. By means of this balance the constraint
determines the scalar $\zeta(x)$ which appears to be generically a
function of $ \overline{\Psi}^{\prime}_{i}\Psi^{\prime}_{i}$.

Existence of a noncanonical contribution to the energy-momentum
tensor (\ref{Tmn-noncan}), along with the $\zeta$ dependence of
the fermion mass (second equation in (\ref{Zeta&m})) means that
generically primordial fermions are very much different from the
regular fermions in the appropriate  field theory model (with the
only volume measure $\sqrt{-g}$). It is however evident that {\em
in normal particle physics conditions}, that is when the energy
density of a single fermion
 is tens of orders of magnitude larger than the
vacuum energy density, the balance dictated by the constraint is
satisfied in the present day universe only if
\begin{equation}
Z_{i}(\zeta)\approx 0 \quad \Longrightarrow \quad
\zeta=\zeta_{1}^{i} \quad or \quad \zeta=\zeta_{2}^{i}, \quad
i=N^{\prime},E^{\prime}. \label{Z-0}
\end{equation}
 with
an extraordinarily high accuracy. Then the fermion mass
(\ref{Zeta&m})  becomes constant and the noncanonical contribution
of the fermion into the energy-momentum tensor,
Eq.(\ref{Tmn-noncan}),  becomes utterly negligible in comparison
with the canonical contribution of the fermion into the
energy-momentum tensor. Thus, if massive fermions are in normal
particle physics conditions, they are described by the standard
field theory equations with constant fermion mass, and, under the
same conditions, gravitational equations (\ref{gef}) with the
energy-momentum tensor determined
 by Eqs.(\ref{Tmn})-(\ref{zeta12})
  are reduced to the Einstein equations in the
appropriate  field theory model (i.e. when the scalar field,
electromagnetic field and massive fermions are sources of
gravity).  Taking into account the undisputed fact that the
classical tests of GR deal only with fermionic matter in normal
particle physics conditions we conclude that although TMT belongs
to alternative gravitational theories, it provides quite standard
picture which includes GR and regular particle physics in all
cases when the fermion energy density
 is tens of orders of magnitude larger than the
vacuum energy density. One can suggest  two alternative approaches
to the question of how a primordial fermion can be realized as a
regular one (see details in Sec.6.2 of Ref.\cite{GK7}).

\subsubsection{Resolution of the 5-th force problem for regular
fermions}

Reproducing Einstein equations when the primordial fermions are in
the states of the regular fermions is not enough in order to
assert that GR is reproduced. The reason is that at the late
universe, as $\phi\gg M_{p}$, the scalar field $\phi$ effective
potential is very flat and therefore due to the Yukawa-type
coupling of massive fermions to $\phi$, (the r.h.s. of
Eq.(\ref{phief+ferm1})),  the long range scalar force appears to
be possible in general. The Yukawa coupling "constant" is
$\alpha\frac{m_{i}(\zeta)}{M_{p}}Z_{i}(\zeta)$. Applying our
analysis of the meaning of the constraint, it is easy to see that
for regular fermions with $\zeta^{(i)} =\zeta_{1,2}^{(i)}$ the
factor $Z_{i}(\zeta)$ is of the order of the ratio of the vacuum
energy density to the regular fermion energy density. Thus we
conclude that the 5-th force is extremely suppressed for the
fermionic matter observable in classical tests of GR. It is very
important that this result is obtained automatically, without
tuning of the parameters and it takes place for both approaches to
realization of the regular fermions in TMT.

\subsection{Nonrelativistic neutrinos and dark energy}

It turns out that besides the normal particle physics situations,
TMT predicts possibility of so called neutrino dark
energy\cite{Nelson}. Roughly speaking such exotic state may be
created in TMT if the degree of localization of the fermion is
very small.  A possible way to get up such a  state might be
spreading of the wave packet of non-relativistic neutrino lasting
a very long (of the cosmological scale) time.

We have studied in detail a model\cite{GK7} where the spatially
flat FRW universe filled with the homogeneous scalar field $\phi$
and  a cold gas of uniformly distributed non-relativistic
neutrinos. Then the constraint (\ref{constraint}) takes the form
\begin{equation}
\frac{(b-\zeta )\left[M^{4}e^{-2\alpha\phi/M_{p}}+
V_{1}\right]-2V_{2}}{(\zeta +b)^{3/2}}=\frac{(\zeta
-\zeta^{(N)}_{1})(\zeta -\zeta^{(N)}_{2})}{(\zeta +k)^{2}}\mu_{N}
\frac{const}{a^{3}}, \label{constraint-toy}
\end{equation}
where we have used that $\overline{N}^{\prime}N^{\prime}=
\frac{const}{a^{3}}$ and $a=a(t)$ is the scale factor.

There is a solution where the decaying fermion contribution  to
the constraint $\sim \frac{const}{a^{3}}$ is  accompanied by
approaching $\zeta\rightarrow -k$ in such a way that $(\zeta
+k)^{-2} \propto a^{3}$. Then both the r.h.s.  and the l.h.s. of
the constraint (\ref{constraint-toy})  approach a constant if
$\phi\rightarrow\infty$ as $a(t)\rightarrow\infty$. The effective
mass of the neutrino in this state increases like $(\zeta
+k)^{-1}\propto a^{3/2}$ and therefore
$\rho_{(N,canon)}=T_{00}^{(N,canon)}=
m_{N}(\zeta)\overline{N}^{\prime}N^{\prime}\propto a^{-3/2}$.
 At the same time   $\Lambda_{dyn}^{(N)}\propto(\zeta
+k)^{-2}\overline{N}^{\prime}N^{\prime}\rightarrow constant$. This
means that at the late time universe, the canonical neutrino
energy density $\rho_{(N,canon)}$
 becomes negligible in comparison with the non-canonical neutrino
 energy density $\rho_{(N,noncanon)}=T_{00}^{(N,noncanon)}=
-\Lambda_{dyn}^{(N)}$. It follows then from Eq.(\ref{Tmn-noncan})
 that such cold neutrino matter possesses a pressure
$p_{N}$ and its equation of state  in the late time universe
approaches the form $p_{N}=-\rho_{N}$ typical for a cosmological
constant. Therefore {\it the primordial non-relativistic neutrino
in the described regime behaves as a sort of dark energy} and
contributions of the quintessence-like field $\phi$ and the
primordial neutrino into the dark energy density are of the same
order of magnitude. We  refer to this regime as
Cosmo-Low-Energy-Physics (CLEP) state. The surprise is that such a
CLEP state is energetically more preferable\cite{GK7} than the one
in the absence of fermions case.

For a particular value of the parameter  $\alpha =\sqrt{3/8}$.
 the cosmological equations allow the following analytic solution\cite{GK7}
 for the late time universe in the CLEP regime:
 \begin{equation}
\phi(t)=\text{const.}+ \frac{M_{p}}{2\alpha}\ln(M_{p}t), \qquad
a(t)\propto t^{1/3}e^{\lambda t}, \label{a-sol-nu}
\end{equation}
where
\begin{equation}
\lambda =\frac{1}{M_{p}}\sqrt{\frac{\Lambda}{3}}, \qquad \Lambda =
\frac{V_{2}-kV_{1}}{(b-k)^{2}}. \label{phi-0}
\end{equation}
 The mass of
the neutrino in such CLEP state increases exponentially in time
and its $\phi$ dependence is double-exponential:
\begin{equation}
m_{N}|_{clep}\sim (\zeta +k)^{-1}\sim a^{3/2}(t)\sim
t^{1/2}e^{\frac{3}{2}\lambda t}\sim \exp\left[\frac{3\lambda
e^{-\varphi_{0}}}{2M_{p}} \exp\left(\frac{2\alpha}{M_{p}}\phi
\right)\right]. \label{m-t-phi}
\end{equation}

\subsection{Prediction of strong gravity effect in high energy
physics experiments}

For the solutions $\zeta\approx\zeta_1$ or $\zeta\approx\zeta_2$
of the constraint (\ref{constraint}), the l.h.s. of the constraint
has the order of magnitude close to the vacuum energy density.
There exists however another solution if one to allow a
possibility that in the core of the support of the fermion wave
function the local dark energy density may be much bigger than the
vacuum energy density. Such a solution turns out to be possible as
fermion density is very big and $\zeta$ becomes negative and close
enough to the value $\zeta\approx -b$. Then the solution of the
constraint (\ref{constraint}) looks\cite{GK5}
\begin{equation}
\frac{1}{\sqrt{\zeta
+b}}\approx\left[\frac{\mu(b-h)}{4M^4b(b-k)}\bar{\Psi}\Psi
e^{2\alpha\phi/M_p}\right]^{1/3}.
 \label{zeta-b}
\end{equation}
In such a case, instead of constant masses, as it was for
$\zeta\approx\zeta_{1,2}$, the second equation in (\ref{Zeta&m})
results in the following fermion self-interaction term in the
effective fermion Lagrangian
\begin{equation}
L^{ferm}_{selfint}=3\left[\frac{1}{b}\left(\frac{\mu(b-h)}{4M(b-k)}\bar{\Psi}\Psi\right)^4
e^{2\alpha\phi/M_p}\right]^{1/3}.
 \label{selfinter}
\end{equation}

It is very interesting that the described effect is the direct
consequence of the strong gravity. In fact, in the regime where
$\zeta +b\ll 1$  the effective Newton constant in the
gravitational term of underlying action(\ref{totaction})
\begin{equation}
S_{grav}=-\int d^{4}x \sqrt{-g}\,\frac{\zeta +b}{\kappa b}
R(\Gamma ,g)e^{\alpha\phi /M_{p}}
 \label{strong-grav}
\end{equation}
becomes anomalously  large. Recall that for simplicity we have
chosen here $b_{\phi}=b_g=b$. But if one do not to imply this fine
tuning then one can immediately see from
Eqs.(\ref{gef})-(\ref{Veff2}) that in the Einstein frame the
regime of the strong gravity dictated by the dense fermion matter
is manifested for the dilaton too.

 The coupling constant in Eq.(\ref{selfinter}) is dimensionless
and depends exponentially of the dilaton $\phi$ if one can regard
$\phi$ as a background field $\phi=\bar{\phi}$. But in a more
general case Eq.(\ref{selfinter}) may be treated as describing an
anomalous dilaton-to-fermion interaction very much different from
the discussed above case of interaction of the dilaton to the
fermion matter in normal conditions where the coupling constant
practically vanishes. Such an anomalous dilaton-to-fermion
interaction should result in  creation of quanta of the dilaton
field in processes with very heavy fermions.  The probability of
these processes is of course proportional to the Newton constant
$M_p^{-2}$. But the new effect consists of the fact that the
effective coupling constant of the anomalous dilaton-to-fermion
interaction is proportional to $e^{2\alpha\bar{\phi}/3M_p}$. If
the dilaton is the scalar field responsible for the quintessential
inflation type of the cosmological scenario\cite{Quint-infl} then
one should expect an exponential amplification of the effective
coupling of this interaction in the present day universe in
comparison with the early universe. One can hope that the
described effect of the strong gravity
 might be revealed  in the LHC experiments in the form of missing energy
 due to the multiple production of quanta of the dilaton
field (recall that coupling of the dilaton to fermions in normal
conditions practically vanishes and therefore the dilaton will not
be observed after  being emitted).

\section{Dust in normal conditions and its decoupling from the dark
energy }

It turns out that the main results of TMT concerning the
decoupling and the restoration of the Einstein's GR in the scale
invariant model \cite{GK5},\cite{GK7} involving fermions and
discussed in the previous section, remain also true in a
macroscopic description of matter.  In addition to the dilaton
$\phi$ dynamics studied in Sec.5, our underlying model now
involves dust as a phenomenological matter model:
\begin{eqnarray}
S&=&S_g+S_{\phi}+S_m \quad \text{where}\label{totaction-1-1}\\
S_{m}&=&\int (\Phi +b_{m}\sqrt{-g})L_m d^{4}x
\label{totaction-1-2}\\
L_m&=&-m\sum_{i}\int e^{\frac{1}{2}\alpha\phi/M_{p}}
\sqrt{g_{\alpha\beta}\frac{dx_i^{\alpha}}{d\lambda}\frac{dx_i^{\beta}}{d\lambda}}\,
\frac{\delta^4(x-x_i(\lambda))}{\sqrt{-g}}d\lambda \label{Lm}
\end{eqnarray}
and $S_g$, $S_{\phi}$ are defined in Eq.(\ref{totaction}) where
again we take for simplicity $b_g=b_{\phi}$. We assume that the
dimensionless parameters $b_g$ and $b_m$ are positive and
 have the same or very close orders of magnitude
 $b_g\sim  b_m$ and besides  $b_m>b_g$. The matter Lagrangian
(\ref{Lm}) describes collection of particles with the same mass
parameter $m$; $\lambda$ is an arbitrary parameter. The model
possesses the global scale invariance (\ref{st}) and we assume in
addition that $x_i(\lambda)$ do not participate in the scale
transformations.

We restrict ourselves to a zero temperature gas of particles, i.e.
we will assume that $d\vec{x}_i/d\lambda \equiv 0$  for all
particles. It is convenient to proceed in the frame where
$g_{0l}=0$, \, $l=1,2,3$. Then the particle density is defined by
\begin{equation}
n(\vec{x})=\sum_{i}\frac{1}{\sqrt{-g_{(3)}}}\delta^{(3)}(\vec{x}-\vec{x}_i(\lambda))
\label{n(x)}
\end{equation}
where $g_{(3)}=\det(g_{kl})$ and
\begin{equation}
S_{m}=-m\int d^{4}x(\Phi
+b_{m}\sqrt{-g})\,n(\vec{x})\,e^{\frac{1}{2}\alpha\phi/M_{p}}
\label{S-n(x)}
\end{equation}

Following the standard procedure  described in Sec.5.1, including
a spontaneous breakdown of the global scale symmetry (\ref{st}) by
means of Eq.(\ref{varphi}) and transition  to the new metric
(\ref{ct}), we obtain the gravitational equations (\ref{gef}) with
the following energy-momentum tensor
\begin{equation}
T_{00}^{eff}=\dot{\phi}^2- \tilde{g}_{00}X
+\tilde{g}_{00}\left[V_{eff}(\phi;\zeta,M) +\frac{3\zeta
+b_m +2b_g}{2\sqrt{\zeta +b_{g}}}\, m\, \tilde{n}\right] \label{T00}\\
\end{equation}
\begin{equation}
T_{ij}^{eff}=\phi_{,k}\phi_{,l}-\tilde{g}_{kl}X
+\tilde{g}_{kl}\left[V_{eff}(\phi;\zeta,M)+\frac{\zeta -b_m
+2b_g}{2\sqrt{\zeta +b_{g}}}\, m\, \tilde{n}\right]\label{Tkl}
\end{equation}
where
\begin{equation}
\tilde{n}(\vec{x})=(\zeta +b_g)^{-3/2}\,
e^{-\frac{3}{2}\alpha\phi/M_{p}}\, n(\vec{x}) \label{ntilde}
\end{equation}
is the particle density in the Einstein frame while the dilaton
$\phi$ field equation  reads
\begin{eqnarray}
\frac{1}{\sqrt{-\tilde{g}}}\partial_{\mu}\left[\sqrt{-\tilde{g}}\tilde{g}^{\mu\nu}\partial_{\nu}\phi\right]
 -\frac{\alpha}{M_{p}(\zeta +b_{g})^{2}} \left[(\zeta
+b_{g})M^{4}e^{-2\alpha\phi/M_{p}}-(\zeta -b_{g})V_{1}
-2V_{2}\right] \nonumber\\
=\frac{\alpha}{M_{p}}\,\frac{\zeta -b_m +2b_g}{2\sqrt{\zeta
+b_{g}}}\, m\,\tilde{n}
 \label{phief-1}
\end{eqnarray}
The scalar field $\zeta$  is determined as a function
$\zeta(\phi,\tilde{n})$ by means of the constraint
\begin{equation}
\frac{(b_{g}-\zeta)\left(M^{4}e^{-2\alpha\phi/M_{p}}+V_{1}\right)-2V_{2}}{(\zeta
+b_g)^2} = \frac{\zeta -b_m +2b_g}{2\sqrt{\zeta +b_{g}}}\, m\,
\tilde{n} \label{constraint2-1}
\end{equation}

One can immediately see the surprising coincidence very similar to
that we noticed in the case of fermions (see paragraph after
Eq.(\ref{constraint})): the explicit $\tilde{n}$ dependence
involving {\em the  same form of $\zeta$ dependence}
\begin{equation}
 \frac{\zeta -b_m +2b_g}{2\sqrt{\zeta +b_{g}}}\, m\,
\tilde{n} \label{universality}
\end{equation}
 appears simultaneously in the noncanonical dust contribution to the pressure (through
the last term in Eq. (\ref{Tkl})), in the  effective dilaton to
dust coupling (in the r.h.s. of Eq. (\ref{phief-1})) and in the
r.h.s. of the constraint (\ref{constraint2-1}). Let us analyze
consequences of this wonderful result in the case when the matter
energy density (modeled by
 dust) is much larger than the dilaton contribution to the dark
 energy density in the space region occupied by this matter. Evidently
this is the condition under which all classical tests of
Einstein's GR, including the question of the fifth force, are
fulfilled. Therefore if this condition is satisfied we will say
that the matter is in {\em normal conditions}. The existence of
the fifth force turns into a problem just in normal conditions.

The last terms in eqs. (\ref{T00}) and (\ref{Tkl}),
 being the matter contributions to the energy density ($\rho_m$) and the
 pressure ($-p_m$) respectively, generally speaking have the same
 order of magnitude. But if the dust is in the normal conditions
 there is a possibility to provide the desirable feature of the dust in GR: it
 must be pressureless. This is realized provided that in normal
 conditions (n.c.) the following equality holds with extremely
 high accuracy:
\begin{equation}
 \zeta^{(n.c.)}\approx b_m-2b_g
\label{decoupling-cond}
\end{equation}
 Inserting
(\ref{decoupling-cond}) in the last term of Eq. (\ref{T00}) we
obtain the effective dust energy density in normal conditions
\begin{equation}
 \rho_m^{(n.c.)}=2\sqrt{b_m-b_g} \, m\tilde{n}
\label{rho-m-n.c.}
\end{equation}
Substitution of (\ref{decoupling-cond}) into the rest of the terms
of the components of the energy-momentum tensor (\ref{T00}) and
(\ref{Tkl}) gives the dilaton contribution to the energy density
and pressure of the dark energy which have the orders of magnitude
close to those in the absence of matter case.

Note that Eq. (\ref{decoupling-cond}) is not just a choice to
provide zero dust contribution to the pressure. The detailed
analysis shows that the constraint (\ref{constraint2-1}) describes
a balance between the pressure of the dust in normal conditions on
the one hand and the vacuum energy density on the other hand. This
balance is realized due to the condition (\ref{decoupling-cond}).

Besides reproducing Einstein equations when the scalar field and
dust (in normal conditions) are sources of the gravity, the
condition (\ref{decoupling-cond}) automatically provides a
practical disappearance of the effective dilaton to matter
coupling. A possible way to see this consists in estimation of the
Yukawa type coupling constant in the r.h.s. of Eq. (\ref{phief}).
Using the constraint (\ref{constraint2}) and representing the
particle density in the form $\tilde{n}\approx N/\upsilon$ where
$N$ is the number of particles in a volume $\upsilon$, one can
make the following estimation for the effective dilaton to matter
coupling "constant" $f$ defined by the Yukawa type interaction
term $f\tilde{n}\phi$ (if we were to invent an effective action
whose variation with respect to $\phi$ would result in Eq.
(\ref{phief})):
\begin{equation}
f \equiv\alpha\frac{m}{M_{p}}\,\frac{\zeta -b_m
+2b_g}{2\sqrt{\zeta +b_{g}}}\approx
\alpha\frac{m}{M_{p}}\,\frac{\zeta -b_m
+2b_g}{2\sqrt{b_m-b_{g}}}\sim
\frac{\alpha}{M_{p}}\,\frac{\rho_{vac}}{\tilde{n}} \approx
\alpha\frac{\rho_{vac}\upsilon}{NM_{p}} \label{Archimed}
\end{equation}
 We conclude that {\it the
effective dilaton to matter coupling "constant" in the normal
conditions is of the order of the ratio of  the "mass of the
vacuum" in the volume occupied by the matter to   the Planck mass
taking $N$ times}. In some sense this result resembles the {\it
Archimedes law}. At the same time Eq. (\ref{Archimed}) gives us an
estimation of the exactness of the condition
(\ref{decoupling-cond}).

Thus our model, formulated both in the microscopic manner (for
fermions) and in the macroscopic manner (for dust), not only
explains why all attempts to discover a scalar force correction to
Newtonian gravity were unsuccessful so far but also predicts that
in the near future there is no chance to detect such corrections.
This prediction is alternative to predictions of other known
models\cite{coupled-quint}-\cite{Alimi}. It is worth here to
stress two crucial distinctions between our approach and models of
Refs.\cite{coupled-quint}-\cite{Alimi}. First, our approach is
based on first principles without introducing into the underlying
action any exotic terms intended to reach desirable results. In
contrast to this, models \cite{coupled-quint}-\cite{Alimi} are
semi-phenomenological. Second, in models based on coupled
quintessence\cite{coupled-quint} and other variable mass particle
ideas\cite{vamp}, the suppression of the effective quintessence to
baryon matter coupling is realized by means of the appropriate
choice (made by hand) of parameters depending of particle species
(or abnormally weighting matter sector in Ref.\cite{Alimi}).
Besides, in the chameleon model\cite{Chameleon}, suppression of
the effective quintessence to baryon matter coupling is achieved
due to a dependence of the scalar field mass upon the matter
density. In our model, the coupling "constant" of the quintessence
to matter depends directly upon the matter density and
automatically, without any special tuning of the parameters, this
coupling "constant" becomes practically zero if matter is in
normal conditions.

Possible cosmological and astrophysical effects when the normal
conditions are not satisfied may be very interesting. In
particular, taking into account that all dark matter known in the
present universe has the macroscopic energy density many orders of
magnitude smaller than the energy density of visible macroscopic
bodies, we hope that the nature of the dark matter can be
understood as a state opposite to the normal conditions.

\bibliographystyle{amsalpha}

\end{document}